# BPS-Saturated Walls in Supersymmetric Theories

B. Chibisov and M. Shifman

*Theoretical Physics Institute, Univ. of Minnesota, Minneapolis, MN 55455*

Domain-wall solutions in four-dimensional supersymmetric field theories with distinct discrete vacuum states lead to the spontaneous breaking of supersymmetry, either completely or partially. We consider in detail the case when the domain walls are the BPS-saturated states, and 1/2 of supersymmetry is preserved. Several useful criteria that relate the preservation of 1/2 of supersymmetry on the domain walls to the central extension appearing in the $N = 1$ superalgebras are established. We explain how the central extension can appear in $N = 1$ supersymmetry and explicitly obtain the central charge in various models: the generalized Wess-Zumino models, and supersymmetric Yang-Mills theories with or without matter. The BPS-saturated domain walls satisfy the first-order differential equations which we call the creek equations, since they formally coincide with the (complexified) equations of motion of an analog high-viscosity fluid on a profile which is given by the superpotential of the original problem. Some possible applications are considered. We also briefly discuss BPS-saturated strings.

# I. INTRODUCTION

Recently a new mechanism for supersymmetry (SUSY) breaking was discussed in [1,2]. The essence of the idea is that higher dimensional theory is compactified on a topological (soliton-like) defect, not invariant under supersymmetry transformations. In the theories where several discrete vacua arise, the domain wall solutions take place. Since each given solution of this type breaks a part of the translational invariance of the theory, it is quite natural that in the sector with the domain wall supersymmetry is spontaneously broken too. In Refs. [1,2] it was observed that in some instances a part of supersymmetry is preserved. The simplest example is the Wess-Zumino model [3]. The domain wall in this model was shown [1] to be the Bogomolny-Prasad-Sommerfeld (BPS) [4] saturated state, thus generalizing the old result [5] referring to kinks in two-dimensional theories. In the present paper we analyze the BPS-saturated domain walls in various models in more detail. The very existence of such states is associated with a central extension of $N = 1$ superalgebra. Since the possibility of such central extensions seemingly contradicts the well-known theorems, we first explain (Sect. 2) what type of central extension can arise, and how the general assertions are circumvented. The key point is that the central extension is automatically zero for all spatially localized field configurations. It need not necessarily vanish, however, for those field configurations that interpolate between distinct vacua at spatial infinities, i.e. in the presence of the domain walls. The concrete form of the centrally extended $N = 1$ superalgebra is derived for the generalized Wess-Zumino models. Here the central charge appears at the classical (tree) level. The central charge in the gauge theories without superpotential emerges at the one-loop level, as a *quantum anomaly*. We give an explicit diagrammatic derivation of this anomaly for supersymmetric QED (SQED). We then comment on how this anomaly is related to those previously established in the non-Abelian theories [6–8] and present the centrally extended $N = 1$ superalgebra for generic non-Abelian theory with matter and superpotential.

In Sect. 3 we return to the generalized Wess-Zumino model, and derive the classical equa-tions defining the BPS-saturated domain walls. These equations turned out to be analogous to those describing the mechanical motion of a high viscosity fluid on a multidimensional profile with mountain ridges and valleys (in the minimal Wess-Zumino model the profile is two-dimensional). A rich physical intuition one has in mechanical motion helps us establish conditions for the existence of the solution(s). If the theory has more than one discrete vacuum state the domain wall as the solution of the classical equations of motion always exists. The BPS saturation equations are stronger constraints, which may or may not have solutions. We establish several useful criteria for the solutions to exist.

The last part of this section is devoted to BPS-saturated strings. The central charge need not vanish in the case of string-like topological defects. If it is non-zero, there arises a possibility of having stable string-like solutions preserving 1/2 of SUSY. We address some general aspects of the BPS-saturated strings and then consider a specific example, in a generalized Wess-Zumino model.

Section 4 is devoted to the minimal Wess-Zumino model. After the domain wall is formed one can view the original $(3+1)$ dimensional theory as a low-energy $(2+1)$ dimensional reduction. We build the reduction explicitly. All fields of the original theory are decomposed in the modes in the wall background. A special role belongs to zero modes – the only ones surviving in the limit of vanishing energy. These modes are localized on the wall. Among the non-zero modes some are localized, others are not. We construct $(2+1)$ dimensional superfields which realize the residual $N=1$ SUSY in three dimensions (this is $N=\frac{1}{2}$ SUSY, from the point of view of the original four-dimensional theory). We obtain a Lagrangian with an infinite number of interacting superfields, in $(2+1)$ dimensions. The resulting theory is explicitly supersymmetric and thus have vanishing corrections to the vacuum energy density. In terms of the four-dimensional theory this means that the domain wall energy density is *not renormalized*. The non-renormalization theorem for the energy density $\varepsilon$ in the BPS-saturated domain walls was established in [1,2]. This phenomenon is related to the existence of the central extension of $N=1$ superalgebra, much in the same way as the exact expression for the monopole mass in the $N=2$ SUSY Yang- Mills theory can be traced back to the central charge appearing in this theory [5].

The representation of $(2+1)$-dimensional supersymmetry is minimal, since there is no analog of the chiral supertransformation in $(2+1)$ dimensions. Unlike the situation in four dimensions, superspace can not be split into two chiral subspaces. This circumstance explains why there are no non-renormalization theorems for superpotential of the type we are used to in four dimensions [9]. As a result, in the $(2+1)$ dimensional theory with $N=1$ SUSY, the shape of the domain wall is subject to perturbative corrections.

In Sect. 4 we also calculate the one-loop perturbative correction to the profile of the domain wall in the minimal Wess-Zumino model.

Section 5 is devoted to sample applications of the ideas developed in this work. First we consider the tunneling problem in multidimensional (non-supersymmetric) quantum mechanics with two or more degenerate classical minima. For a class of potentials of a special form the tunneling trajectory (in the Euclidean time) is determined by the same first-order equations as the domain wall profiles in the generalized Wess-Zumino models. The fact that these equations do have solutions can be established from the global features of the potentials, without actual solutions. The analytic expression for the classical action on these least-action trajectories follows immediately. This classical action determines the tunneling exponent.

In the second part of Sect. 5 we make use of the calculated quantum corrections to the profile of the domain wall. They are related to the subleading corrections in the asymptotic behavior of the multiparticle production amplitudes at threshold when the number of particles tends to infinity [10,11]. By exploiting this relation we find the first subleading correction to the corresponding asymptotic formula in the minimal Wess-Zumino model. This correction was previously known in non-supersymmetric theories [11,13,12].

In Sect. 6 we address the issue of renormalization of the soliton mass $\mu$ in the Witten-Olive solution of the $(1+1)$ dimensional theory. The reason why $\mu$ is renormalized while $\varepsilon$ in four-dimensional theories is not, is revealed.

Technical details of the formalism extensively exploited throughout the paper are col-lected in Appendices.

Some of the results referring to the generalized Wess-Zumino model, to be considered below, were discussed previously, mainly in the stringy context, see [15,16] and references therein. In particular, the arguments at the end of Sect. 3.A can be found in these works. We present them for the sake of completeness. Our main emphasis is put on new results: (i) the anomalous term due to the gauge fields appearing in the central extension of $N = 1$ superalgebra; (ii) criteria for the existence of the BPS-saturated domain walls which is far from being guaranteed by the occurrence of a non-vanishing central extension; (iii) so far unexplored applications.

## II. CENTRAL EXTENSIONS

### A. Generalities

In this section we explain how and what kind of central extension can appear in $N = 1$ superalgebra

$$\{Q_\alpha Q_\beta\} = 2T_{\alpha\beta}, \qquad (1)$$

where $Q_\alpha$ are the supercharges, $\alpha = 1, 2$ (the Hermitian conjugated supercharges will be denoted $\bar{Q}_{\dot{\alpha}}$). It is common wisdom, that $N = 1$ supersymmetric theories can not contain central extensions. The general classification of superalgebras dates back to the classical paper [17]. Should the central extension appear in the anticommutator (1), it will clearly belong to $(0, 1)$ representation of the Lorentz group. Here $Q_\alpha$ is the supercharge, and spinorial notation is used throughout the paper, if not stated to the contrary. This fact is obvious, since the central extension should be symmetric with respect to two undotted indices. The existence of an extra conserved quantity that is not a Lorentz scalar, in addition to four-momentum, is forbidden by the Coleman-Mandula theorem [18] for all Lorentz-invariant theories with non-trivial $S$ matrix. Thus, in order to have a central extension we must violate some assumptions of this theorem. Let us look at the theories where the Lorentz symmetry is spontaneously broken. This situation can be realized in the theories with spontaneously broken discrete symmetries, with two or more discrete degenerate vacua. Field configurations that interpolate between the distinct vacua are extended objects which are not invariant under the action of the Poincaré group. If we choose such a field configuration as our "vacuum", then we are clearly in the situation with the spontaneously broken Lorentz symmetry. In this case the existence of a central extension does not contradict the Coleman-Mandula theorem, $T_{\alpha\beta} \neq 0$. The central charge, however, must vanish in the sector with the Lorentz-invariant vacuum.

In supersymmetric theories one can have a discrete set of (degenerate) vacua even without spontaneous breaking of any discrete symmetry. Supersymmetry, rather than the spontaneous breaking of the discrete symmetry, will keep the energy of these states at zero, so that the interpolating field configurations, the domain walls, can be stable both classically and quantum-mechanically. As a matter of fact, this situation is quite typical. Supersymmetric gluodynamics, the simplest SUSY gauge theory, does have $N+1$ vacuum states (for the gauge group $SU(N)$) that are not continuously connected. $N$ vacua are chirally asymmetric and are related to the spontaneous breaking of the $Z_{2N}$ symmetry; one extra vacuum is a recently discovered [19] chirally symmetric state whose existence is unrelated to the spontaneous breaking of any discrete symmetry. In the generalized Wess-Zumino models multiple vacuum states unrelated to the spontaneous breaking of any discrete symmetry are quite conventional too. The BPS-saturated domain walls are abundant in these theories, the phenomenon is not at all exotic.

The supercharges appearing in the superalgebra (1) are defined as

$$Q_\alpha = \int d^3x J^0_\alpha. \tag{2}$$

where $J^\mu_\alpha$ is the conserved supercurrent of the supersymmetric model at hand, in the spin-vector form,

$$J^\mu_\alpha = \frac{1}{2}(\bar\sigma^\mu)^{\dot\beta\beta} J_{\alpha\beta\dot\beta}, \tag{3}$$

where $J_{\alpha\beta\dot{\beta}}$ is the supercurrent in the spinorial form, and the matrices $\bar{\sigma}$ are defined in Appendix A, where all our notations and conventions are collected. The current conservation can be written as

$$\partial_\mu J_\alpha^\mu \equiv \frac{1}{2}\partial^{\dot{\beta}\beta} J_{\alpha\beta\dot{\beta}} = 0. \tag{4}$$

The supercurrent $J_{\alpha\beta\dot{\beta}}$ is a member of the supercurrent supermultiplet

$$J_{\alpha\dot{\alpha}} = R^0_{\alpha\dot{\alpha}} - \frac{1}{2}\left\{i\theta^\beta(J_{\beta\alpha\dot{\alpha}} - \frac{2}{3}\epsilon_{\beta\alpha}\epsilon^{\gamma\delta}J_{\delta\gamma\dot{\alpha}}) + H.C.\right\} -$$

$$\theta^\beta\bar{\theta}^{\dot{\beta}}\left(J_{\alpha\dot{\alpha}\beta\dot{\beta}} - \frac{1}{3}\epsilon_{\alpha\beta}\epsilon_{\dot{\alpha}\dot{\beta}}\epsilon^{\gamma\delta}\epsilon^{\dot{\gamma}\dot{\delta}}J_{\gamma\dot{\gamma}\delta\dot{\delta}}\right) + \ldots \tag{5}$$

where $R^0_{\alpha\dot{\alpha}}$ is the $R_0$ current [20] and $J_{\alpha\dot{\alpha}\beta\dot{\beta}}$ is the energy-momentum tensor [6]. The relation between $\theta_{\mu\nu}$ and $J_{\alpha\dot{\alpha}\beta\dot{\beta}}$ is given in Appendix A, Eq.(A41).

With these definitions in hands we can proceed now to a detailed discussion of the issue how $T_{\alpha\beta} \neq 0$ appears.

### B. Wess-Zumino Model

In the non-gauge theories, the presence of the central extension can be seen at the tree level. The most clear-cut example is the Wess-Zumino model, where we can explicitly find the solution of the classical equations of motion interpolating between two vacua of the model (the domain wall). In this case the central extension appears already at the tree level. The Wess-Zumino Lagrangian in terms of superfields has the following form,

$$\mathcal{L} = \frac{1}{4}\int d^4\theta \Phi\bar{\Phi} + \left\{\frac{1}{2}\int d^2\theta \mathcal{W}(\Phi) + H.C.\right\}, \tag{6}$$

$$\mathcal{W}(\Phi) = \mu^2\,\Phi - \frac{\lambda}{3}\Phi^3, \tag{7}$$

where the parameters $\mu$ and $\lambda$ can be always chosen to be real. If $\mu = 0$, the Lagrangian (6) is invariant under the $R_0$-rotations [20],

$$\theta \to e^{i\alpha}\theta\,, \quad \phi \to e^{\frac{2i\alpha}{3}}\phi \tag{8}$$

at the classical level. If $\mu \neq 0$ this invariance is gone, but a discrete $Z_2$ subgroup persists,

$$\phi \to -\phi, \quad \psi \to i\psi, \quad F \to F. \tag{9}$$

This discrete part of $R_0$ symmetry does not mix the real and imaginary parts of the bosonic field. The $Z_2$ symmetry is spontaneously broken, and the corresponding domain wall solution (see Sect. 4), that interpolates between two degenerate vacua, exists. The supercurrent has the following form (we write everything in the chiral notation, see Appendix A),

$$J_{\alpha\beta\dot{\beta}} = 2\sqrt{2} \left\{ \left[ (\partial_{\alpha\dot{\beta}}\phi^+)\psi_\beta - i\,\epsilon_{\beta\alpha}F\bar{\psi}_{\dot{\beta}} \right] - \frac{1}{6} \left[ \partial_{\alpha\dot{\beta}}(\psi_\beta\phi^+) + \partial_{\beta\dot{\beta}}(\psi_\alpha\phi^+) - 3\epsilon_{\beta\alpha}\partial^\gamma_{\dot{\beta}}(\psi_\gamma\phi^+) \right] \right\}. \tag{10}$$

Its dependence on the superpotential comes through the equation of motion for the $F$ term,

$$F = -\frac{\partial \bar{\mathcal{W}}}{\partial \phi^+}. \tag{11}$$

The second square brackets in Eq. (10) contain a full spatial derivative and, in principle, could have been omitted. Upon inspecting Eq.(10) it becomes clear that the anticommutator of two supercharges does not vanish, and is completely determined by the canonic equal-time anticommutator

$$\{\psi_\alpha(x)\bar{\psi}_{\dot{\alpha}}(y)\}_{\text{e.t.}} = \mathbf{1}_{\alpha\dot{\alpha}}\delta(\vec{x}-\vec{y}). \tag{12}$$

The term with $\partial_0\phi^+(\sigma_0)_{\beta\dot{\alpha}}$ drops out since it can not yield symmetric in $(\alpha,\beta)$ structures, and then we get for the anticommutator of two supercharges,

$$\{Q_\alpha\,Q_\beta\} = (-4i)(\vec{\sigma})_{\alpha\beta}\int d^3x\vec{\nabla}\left\{\bar{\mathcal{W}} - \frac{1}{3}\bar{\Phi}\frac{\partial\bar{\mathcal{W}}}{\partial\bar{\Phi}}\right\}_{\bar{\theta}=0}. \tag{13}$$

The matrix $\vec{\sigma}_{\alpha\beta}$ is defined in Eq.(A12) and is automatically symmetric in $\alpha, \beta$. It converts the spinorial indices of the representation (1,0) of the Lorentz group in the vectorial form. The expression on the right-hand side is formally ambiguous. Indeed, let us drop the second term in the supercurrent (10). This term is a full spatial derivative and should have no impact on the supercharge. However, if we calculated $\{Q_\alpha Q_\beta\}$, using only the first term in Eq.(10) for defining the supercharge $Q_\alpha$, we would get

$$\{Q_\alpha\, Q_\beta\} = (-4i)(\vec{\sigma})_{\alpha\beta} \int d^3x\, \vec{\nabla}\bar{\mathcal{W}}\,|_{\bar{\theta}=0}\,, \tag{14}$$

instead of Eq.(13). This ambiguity does *not* affect the value of the central charge, however. The term $\bar{\Phi}\partial\bar{\mathcal{W}}/\partial\bar{\Phi}$ is a full superderivative, and it vanishes in any supersymmetric vacuum. Thus, whatever wall is considered, $\bar{\Phi}\partial\bar{\mathcal{W}}/\partial\bar{\Phi}$ gives no contribution in Eq.(13).

We pause here to make a remark regarding the impact of loop corrections in Eq. (13). Strictly speaking, when one includes the quantum loops, the tree-level result (13) acquires an additional term on the right-hand side, due to the quantum anomaly. This term is

$$\frac{1}{8}\gamma D^2(\bar{\Phi}\Phi)$$

where $\gamma$ is the anomalous dimension of the superfield $\Phi$, cf. the last term in Eq. (40). As we already know, such terms (full superderivatives) can be freely omitted.

Summarizing, the expression for the central charge given in Eq.(14) is *exact*. The operator on the right-hand side, being sandwiched between any states in the sector with the unbroken Lorentz symmetry (no domain walls or other topological defects) vanishes, in accordance with the Coleman-Mandula theorem. In the sector with the domain wall, however, the expression is non-zero. It is saturated by the extended field configuration of the domain wall and is proportional to the difference between the vacuum expectation values of $\mathcal{W}$ in two distinct vacua, between which the given domain wall interpolates.

In the example considered the value of $\mathcal{W}$ in both vacuum states is real. This is not necessarily the generic case. If one limits oneself to consideration of the given pair of vacua (or the model considered has exactly two vacuum states) it is always possible to ensure that $\mathcal{W}$ in both vacuum states is real. Indeed, first we can use the fact that the superpotential is defined up to an additive constant. Using this freedom one can always choose $\mathcal{W}_{*1} = 0$. Here the subscript $*1$ marks the value of the superpotential in the first vacuum. Moreover, the superpotential is defined up to a phase. The phase of $\mathcal{W}$ can be always rotated by an appropriate rotation of the Grassmann variable $\theta$. Using this freedom it is always possible to make $\mathcal{W}_{*2}$ real. It is even possible to make $\mathcal{W}_{*2}$ real and positive. Below this choice will be referred to as *standard*, and in many instances we will exploit it. (Of course, if one discusses simultaneously two or more distinct domain walls in the problems with three or more vacua it may not be always possible to kill the phase.)

It is convenient to define the central charge under discussion in the following general form. Let the charges be normalized in the standard way,

$$\{Q_\alpha \bar{Q}_{\dot{\beta}}\} = 2P_{\alpha\dot{\beta}}, \tag{15}$$

and assume that the wall lies in the $xy$ plane (this can always be achieved by an appropriate choice of the reference frame). If

$$\{Q_\alpha Q_\beta\} = (-2i)(\tau_1)_{\alpha\beta}\bar{\Sigma}A, \tag{16}$$

the parameter $\Sigma$ will be referred to as the central charge. Here $A$ is the area of the wall. From the consideration above it is clear that in the Wess-Zumino model

$$\bar{\Sigma} = 2\left[\bar{\mathcal{W}}(z = +\infty) - \bar{\mathcal{W}}(z = -\infty)\right]. \tag{17}$$

For the superpotential (7) the vacua are at

$$\phi_{*1,2} = \pm\frac{\mu}{\sqrt{\lambda}}$$

and

$$\Sigma = \frac{8}{3}\frac{\mu^3}{\sqrt{\lambda}}.$$

Although the derivation was given for the minimal Wess-Zumino model, Eqs. (14) and (17) are obviously valid for *any number of the chiral superfields and arbitrary superpotential*.

Since the supercharges are conserved, the right-hand side of Eq.(16) is a conserved quantity too. This observation can be viewed as an alternative proof of the absence of renormalization of the superpotential. It is well-known [9] that superpotentials are not renormalized perturbatively. Non-perturbative renormalization can occur only if Eq.(17) has a non-trivial anomaly (see below).

The commutator relation (16) contains only $Q$'s not $\bar{Q}$'s. This means that the central charge $\bar{\Sigma}$ must be an analytical function of various parameters characterizing the problem (masses, coupling constants, and so on).

The fact that we were able to find an expression for the central extension as a local operator does not necessarily mean that the central charge is non-vanishing. For instance, if the parameter $\mu$ in Eq.(7) is put to zero, $\Sigma = 0$. In Ref. [1] examples are given of the generalized Wess-Zumino models with two classically distinct vacua such that

$$\mathcal{W}(z = +\infty) = \mathcal{W}(z = -\infty). \tag{18}$$

In these models $\Sigma$ vanishes too. The domain walls are still present, of course, but they are not BPS-saturated.

Assume now that $\Sigma \neq 0$. By combining Eqs. (15) and (16) we can get a bound for the domain wall energy density $\varepsilon$.

<u>*Theorem 1.*</u>   $\varepsilon \geq |\Sigma|$. If the inequality is saturated, i.e. $\varepsilon = |\Sigma|$, 1/2 of SUSY is unbroken on the given domain wall.

To prove the theorem consider a linear combination of operators $Q$ and $\bar{Q}$,

$$K = \beta^\alpha Q_\alpha + \bar{\beta}^{\dot{\alpha}} \bar{Q}_{\dot{\alpha}} \tag{19}$$

where $\beta$ is an arbitrary complex parameter, with two components, $\beta = \{\beta^1, \beta^2\}$. Both are treated as $c$-numbers rather than the Grassmann numbers. Since $K$ is Hermitian, the expectation value over any state $S$

$$\langle S|K^2|S\rangle \geq 0. \tag{20}$$

Equation (20) implies, in turn, that for any $\beta$

$$\mathcal{E}(|\beta^1|^2 + |\beta^2|^2) + 2\mathrm{Im}\left(\beta^1 \beta^2 \bar{\Sigma}\right) A \geq 0 \tag{21}$$

where $\mathcal{E}$ is the energy of the state $S$. To get the best possible bound on $E$ we must optimize our choice of $\beta$. If

$$\bar{\Sigma} = \rho e^{-i\alpha} \qquad (22)$$

optimization is evidently achieved at $\beta^1 = -i\beta^2 = be^{i\alpha/2}$ where $\rho$ and $b$ are positive numbers, while $\alpha$ is a phase. Equation (22) then reduces to

$$E \geq \rho A \qquad (23)$$

or, in other words,

$$\varepsilon \geq |\Sigma|, \qquad (24)$$

q.e.d. The walls for which this inequality is saturated (i.e $\varepsilon = |\Sigma|$) are called BPS-saturated. It is clear that the BPS saturation can be achieved only provided that a linear combination of the supercharges, acting on the wall, annihilates it. This means, that a part of supersymmetry is preserved in the sector with the given domain wall.

### C. Central charge in $N=1$ gauge theories.

Let us discuss now supersymmetric gauge theories with or without the superpotential. As a warm-up exercise we consider supersymmetric electrodynamics (SQED). The Lagrangian is

$$\mathcal{L} = \left(\frac{1}{8e^2}\int d^2\theta W^2 + H.C.\right) + \frac{1}{4}\int d^4\theta \left(\bar{S}e^V S + \bar{T}e^{-V}T\right) + \left(\frac{m_0}{2}\int d^2\theta S\, T + H.C.\right) \qquad (25)$$

where $S$ and $T$ are chiral superfields of the electric charge $+1$ and $-1$, respectively. The Lagrangian (25) describes photon, electron and two selectrons. The expression for the supercharge can be read off Eq.(A45).

The model possesses the $R_0$ symmetry

$$\lambda \to e^{i\alpha}\lambda, \quad \psi_{s,t} \to e^{-i\alpha/3}\psi_{s,t}, \quad \phi_{s,t} \to e^{2i\alpha/3}\phi_{s,t}, \qquad (26)$$

which is broken down to $Z_2$ by the mass term and the quantum anomaly. The vacuum is non-degenerate and, thus, we do not expect to find the domain walls. Nevertheless, it is still useful to complete the calculation of $T_{\alpha\beta}$. The supercurrent of the model is

$$J_{\alpha\beta\dot\beta} = 2\left[2i\, G_{\beta\alpha}\bar\lambda_{\dot\beta} - 6\epsilon_{\beta\alpha}D\bar\lambda_{\dot\beta} + \sqrt{2}\left\{(\partial_{\alpha\dot\beta}\phi^+)\psi_\beta - i\,\epsilon_{\beta\alpha}F\bar\psi_{\dot\beta}\right\}\right]. \tag{27}$$

Terms with the total spatial derivatives in the supercurrent, see Eq.(A45), are unimportant for the present discussion and we omitted them from the beginning. In the tree approximation the anticommutator $\{Q_\alpha Q_\beta\}$ is the same as in the Wess-Zumino model with

$$\mathcal{W} = m_0 S\, T, \tag{28}$$

since the photon (photino) part obviously gives no contribution to this anticommutator at the classical level. However, the gauge field contribution emerges at the one-loop level due to the quantum anomaly. This is most readily seen by using the Pauli-Villars regularization (Fig.1). If we add the Pauli-Villars fields to render the theory ultraviolet- finite, all commutators do reduce to the naive tree-level expressions, with the regulator term explicitly added. In particular, in SQED

$$\{Q_\alpha\, Q_\beta\} = (-4i)(\vec\sigma)_{\alpha\beta}\int d^3x\,\vec\nabla\left[m(\phi_s^*\phi_t^*) + M_0(R_s^*R_t^*)\right], \tag{29}$$

where $R_{s,t}$ is the lowest component of the regulator chiral superfield. It enters in the Lagrangian in the same way as $S,T$, with the substitution $m \to M_0$ ($M_0 \to \infty$) and the opposite metric, i.e. the sign of the regulator loop is opposite to the normal one.

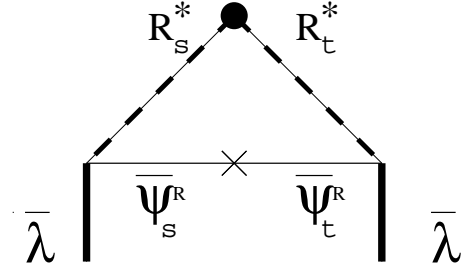

FIG. 1. The anomalous part in the central charge.

Integrating out the regulator field in the gluino background (this amounts to calculating the diagram depicted on Fig. 1) we obtain:

$$\{Q_\alpha\, Q_\beta\} = (-4i)(\vec\sigma)_{\alpha\beta}\int d^3x\,\vec\nabla\left[m(\phi_s^*\phi_t^*) - \frac{1}{16\pi^2}\bar\lambda\bar\lambda\right]. \tag{30}$$

The right-hand side can be written as the lowest component of the superfield in the operator equation,

$$\{Q_\alpha\, Q_\beta\} = (-4i)(\vec{\sigma})_{\alpha\beta} \int d^3x \vec{\nabla} \left[ m\, \bar{S}\, \bar{T} + \frac{1}{16\pi^2} \bar{W}\bar{W} \right]_{\bar{\theta}=0} \tag{31}$$

where $W_\alpha$ is the chiral superfield generalizing the photon field strength tensor. The expression in the square brackets is nothing else but the Konishi anomaly [7] (see appendix A and Eq.(41)). As was expected, the central extension is determined by the quantum anomaly of the theory. Since it is a full superderivative,

$$\frac{1}{8}\bar{D}^2 \left( \bar{S}e^V S + \bar{T}e^{-V} T \right),$$

the central charge vanishes in SQED, and the derivation of Eq. (31) is meaningful only to the extent that it graphically illustrates how the anomaly (the term $\propto W^2$) can appear in $T_{\alpha\beta}$. SQED is clearly uninteresting in this context and is studied only as a technical exercise. Although the central charge is identically zero in this case, we still find it instructive to examine the formal operator appearing on the right-hand side of Eq.(31) under the *given* definition of the supercharges, see Eq.(27).

Let us proceed now to much more interesting case of the non-Abelian theories. In the non-Abelian gauge theories, the direct calculation of the central extension is more complicated, for it is necessary to include the contribution of the gauge fields and, as always, one has to find a way to regularize both, the ultraviolet and infrared divergences in a gauge and superinvariant way. It is easier to express the anomaly in $\{Q_\alpha Q_\beta\}$ in an indirect way by exploiting the anomaly relations already established in the theory.

For definiteness let us consider $SU(N_c)$ SQCD with $N_f$ flavors. Each flavor consists of two subflavors, $Q_\alpha$ and $\tilde{Q}_\alpha$ where $\alpha$ is a $SU(N)$ index. The mass term is assumed to be diagonal,

$$\mathcal{W}_m = \Sigma_f \left( m Q^\alpha \tilde{Q}_\alpha \right)_f \tag{32}$$

where the sum runs over all flavors. Such a diagonalization is always possible. In the general case more generic terms in the superpotential are allowed. Our final result will be valid for a generic superpotential.

The set of "geometric" anomalies of SQCD includes that in the $R_0$ current, supercurrent and the trace of the energy-momentum tensor. In the superfield notation this set is represented by the anomaly in the supermultiplet of currents [6,8],

$$\bar{D}^{\dot\alpha} J_{\alpha\dot\alpha} = \frac{1}{3} D_\alpha \left\{ \left[ 3\mathcal{W} - \sum_i Q_i \frac{\partial \mathcal{W}}{\partial Q_i} \right] - \left[ \frac{3T(G) - \sum_i T(R_i)}{16\pi^2} \text{Tr} W^2 + \frac{1}{8} \sum_i \gamma_i Z_i \bar{D}^2 (\bar{Q}_i e^V Q_i) \right] \right\}. \quad (33)$$

This is a general expression, valid for any gauge theory, with the arbitrary gauge group and superpotential. In SQCD we must use superpotential (32), while $T(G) = N_c$ and $T(R_i) = 1$ for each flavor. Equation (33) is valid even in the presence of the Yukawa (trilinear) couplings in the superpotential.

The lowest component of the current supermultiplet $J_{\alpha\dot\alpha}$ is the $R_0$ current, while the $\theta$ and $\bar\theta$ components of the supermultiplet are related to the supercurrent $J_{\beta\alpha\dot\alpha}$, see Eq. (5). Now, we can obtain the central extension by investigating the commutator of the supercharge $\bar{Q}_{\dot\alpha}$ with the supercurrent,

$$\{\bar{Q}_{\dot\alpha} \bar{Q}_{\dot\beta}\} = \frac{1}{2} (\bar{\sigma}^0)^{\gamma\dot\gamma} \int d^3x \{\bar{Q}_{\dot\alpha} \bar{J}_{\dot\beta\dot\gamma\gamma}\}. \quad (34)$$

With the aid of Eq.(5), the supercurrent can be expressed in terms of the $\bar\theta$ component of the current supermultiplet,

$$\frac{-i}{2} \bar{J}_{\dot\beta\dot\gamma\gamma} = 2\{J_{\gamma\dot\beta}\}_{\bar\theta^{\dot\gamma}} - \{J_{\gamma\dot\gamma}\}_{\bar\theta^{\dot\beta}}. \quad (35)$$

Then we have

$$\frac{1}{2} \{\bar{Q}_{\dot\alpha} \bar{J}_{\dot\beta\dot\gamma\gamma}\} = 2i\{\bar{Q}_{\dot\alpha} J_{\gamma\dot\beta}\}_{\bar\theta^{\dot\gamma}} - i\{\bar{Q}_{\dot\alpha} J_{\gamma\dot\gamma}\}_{\bar\theta^{\dot\beta}}. \quad (36)$$

The $\bar\theta$ component of the anticommutators on the right-hand side can be rewritten as the lowest component of the superderivative of the current,

$$2i\{\bar{Q}_{\dot\alpha} J_{\gamma\dot\beta}\}_{\bar\theta^{\dot\gamma}} - i\{\bar{Q}_{\dot\alpha} J_{\gamma\dot\gamma}\}_{\bar\theta^{\dot\beta}} = \left\{ \epsilon_{\dot\gamma\dot\alpha} \bar{D}^2 J_{\gamma\dot\beta} - \frac{1}{2} \epsilon_{\dot\beta\dot\alpha} \bar{D}^2 J_{\gamma\dot\gamma} \right\}_{\theta=0}. \quad (37)$$

The last term being antisymmetric with respect to the indices $\dot{\alpha}$ $\dot{\beta}$ does not contribute to the anticommutator of the supercharges and we drop it, while for the first term we obtain,

$$\left\{\epsilon_{\dot{\gamma}\dot{\alpha}}\bar{D}^2 J_{\gamma\dot{\beta}}\right\}_{\theta=0} = 2\left\{\epsilon_{\dot{\gamma}\dot{\alpha}}\bar{D}_{\dot{\beta}}\bar{D}^{\dot{\delta}} J_{\gamma\dot{\delta}}\right\}_{\theta=0}. \tag{38}$$

Assembling everything together we get

$$\{\bar{Q}_{\dot{\alpha}}\bar{Q}_{\dot{\beta}}\} = -2(\sigma^0)^{\gamma}_{\dot{\alpha}} \int d^3x \left\{\bar{D}_{\dot{\beta}}\bar{D}^{\dot{\delta}} J_{\gamma\dot{\delta}}\right\}_{\theta=0}. \tag{39}$$

Substituting the anomaly equation (33) into the superderivative of the current we get

$$\{\bar{Q}_{\dot{\alpha}}\bar{Q}_{\dot{\beta}}\} = \frac{-4i}{3}(\vec{\sigma})_{\dot{\alpha}\dot{\beta}} \int d^3x \vec{\nabla} \left\{\left[3\mathcal{W} - \sum_i Q_i \frac{\partial \mathcal{W}}{\partial Q_i}\right] - \left[\frac{3T(G) - \sum_i T(R_i)}{16\pi^2}\text{Tr}W^2 + \frac{1}{8}\sum_i \gamma_i Z_i \bar{D}^2(\bar{Q}_i^+ e^V Q_i)\right]\right\}_{\substack{\bar{\theta}=0 \\ \theta=0}}. \tag{40}$$

This is our master formula. In deriving it we took into account the fact that the term with time derivative $\partial_0$ is proportional to $\epsilon_{\dot{\alpha}\dot{\beta}}$ and cancels out after the symmetrization over the indices $\dot{\alpha}$, $\dot{\beta}$, while those with spatial derivatives are proportional to the matrix $(\vec{\sigma})$

$$(\sigma^0)^{\gamma}_{\dot{\alpha}}(\sigma^i)_{\gamma\dot{\beta}}\partial^i = (\vec{\sigma})_{\dot{\alpha}\dot{\beta}}\vec{\nabla}.$$

Equation (40) is a general, multiloop expression for the central extension of the $N = 1$ superalgebra in the gauge theories with a superpotential. If $\mathcal{W} \neq 0$, as was mentioned above, the superalgebra acquires the central extension already at the tree level. Note that the classical part (one containing the superpotential and its derivatives) in Eq.(40) is in agreement with that obtained for the Wess-Zumino model (Eq.(13)). As was expected, the result is the integral over a total derivative. It can only be supported by extended field configurations. Moreover, Eq.(40) contains total superderivatives that do not contribute to the central charge (i.e. to the matrix elements in the supersymmetric vacua). One such total derivative is explicitly present in Eq.(40), while the other can be seen by invoking again the Konishi anomaly [7],

$$\frac{1}{8}\bar{D}^2(\bar{Q}_i^+ e^V Q_i) = \frac{1}{2}Q_i\frac{\partial \mathcal{W}}{\partial Q_i} + \frac{\sum_i T(R_i)}{16\pi^2}\text{Tr}W^2. \tag{41}$$

Subtracting the Konishi anomaly from the right-hand side of the Eq.(40) we obtain

$$\{\bar{Q}_{\dot\alpha}\bar{Q}_{\dot\beta}\} = (-4i)(\vec{\sigma})_{\dot\alpha\dot\beta} \int d^3x \vec{\nabla} \left\{ \left[ \mathcal{W} - \frac{T(G) - \sum_i T(R_i)}{16\pi^2} \mathrm{Tr} W^2 \right] + \mathrm{t.s.d.} \right\}_{\substack{\bar\theta=0 \\ \theta=0}} \qquad (42)$$

where t.s.d. denotes total superderivatives. The expectation value of the operator appearing on the right-hand side over any supersymmetric state is renormalization-group-invariant, to *all orders*. This was established in Ref. [21], see Sect. 4. This fact can serve as an independent prove that the central extension of $N = 1$ superalgebra derived above for supersymmetric gauge theories is exact to all orders.

We can now compare this result with the independent calculation of the central extension in the Abelian case (Eq.(31)). Indeed, for the Abelian gauge group and superpotential (32) Eq.(42) reduces again to the Konishi anomaly, in accordance with Eq.(31). Unlike SQED, Eq.(42) is non-trivial. The non-Abelian gauge theories generically do possess discrete sets of vacua, giving rise to the possibility of the domain wall solution [2,22]. Correspondingly, the central charge need not vanish. The non-vanishing central charge in SQCD will be transparently seen from the consideration below.

It is quite natural, that the central extension is expressed in terms of anomalies already existing in the theory. If it were not the case, it would imply the existence of a new anomaly of a geometric nature, not known so far, that appears only in the anticommutator of the supercharges. No such new anomaly exists.

To reveal in a more graphic form the relation between the "old" anomalies known in the gauge theories, and the anomalous term in the central charge under consideration, we suggest a slightly different derivation based on the so called Veneziano-Yankielowicz effective Lagrangian [23,24]. The advantage of this derivation is that it translates the (anomalous) loop effects in a straightforward tree calculation similar to that performed in the Wess-Zumino model.

From dimensional counting and the analysis of the $R$ charges, it is clear that the most general expression for the central extension of $N = 1$ superalgebra has the form,

$$\{Q_\alpha Q_\beta\} = (-4i)(\vec{\sigma})_{\alpha\beta} \int d^3x \vec{\nabla} \left[ \bar{\mathcal{W}}(\bar{Q}, \bar{\tilde{Q}}) + C \, \mathrm{Tr} \bar{W}^2 \right]_{\bar\theta=0} \qquad (43)$$

where $C$ is a numerical constant. Dimensional arguments and $R$ symmetry tell us that the anomalous part can be proportional only to the operator $\text{Tr}\bar{W}^2$. There is no other gauge invariant operator in the theory, built from the gluon (super) fields, with the appropriate dimension and $R$ charge. Our task is determining the numerical value of the constant $C$.

The anomaly content of the theory is explicit in the Veneziano-Yankielowicz effective Lagrangians [23,24].

If we introduce the (composite) superfields

$$S = \frac{3}{32\pi^2}\text{Tr}W^2 \;,$$
$$M^i_j = Q^i \tilde{Q}_j \;, \quad i,j = 1,2,\ldots,N_f \;, \tag{44}$$

the effective Lagrangian incorporating all standard anomalies takes the form

$$\mathcal{L} = \frac{1}{4}\int d^4\theta \left(\bar{S}\,S\right)^{1/3} + \frac{1}{4}\int d^4\theta \left(\text{Tr}\bar{M}\,M\right) +$$
$$\left[\frac{1}{3}\int d^2\theta S \ln \frac{S^{N_c-N_f}\det M}{e^{N_c-N_f}} + \frac{1}{2}\int d^2\theta \text{Tr}\,(mM) + H.C.\right] \;, \tag{45}$$

up to corrections discussed in Ref. [19]. These corrections are irrelevant for our present purposes. The scale factor $\Lambda$ is put to unity, so that all dimensionfull quantities are measured in units of $\Lambda$. The specific choice of the kinetic term in the expression above is unimportant. What is important is the normalization of the various terms in the superpotential. We pause here to make a few comments on how this normalization is fixed. First, in the massless limit, SQCD with $N_f$ flavors possesses an anomaly-free $R$ symmetry [25],

$$\theta \to \theta e^{i\alpha}\;,\quad W \to W e^{i\alpha}\;,\quad M^i_j \to M^i_j e^{2i\alpha N_f^{-1}(N_f-N_c)} \tag{46}$$

and, additionally, $SU(N_f)_L \times SU(N_f)_R$ global invariance. The latter requires the effective Lagrangian to depend on $\det M$, while the former unambiguously determines the argument of the logarithm in Eq.(45). The coefficient of the first term in the square brackets can be found by inspecting, for instance, the conformal anomaly. Under the transformations

$$x \to x(1+\delta)\;,\quad G^a_{\mu\nu} \to G^a_{\mu\nu}(1-2\delta)\ldots \tag{47}$$

the action of the fundamental theory

$$\delta \mathcal{A} = \frac{3N_c - N_f}{32\pi^2} \int G^a_{\mu\nu} G^a_{\mu\nu} d^4 x. \tag{48}$$

The effective action (45) has the same variation. Since it is supersymmetric, all other "geometric" anomalies are automatically reproduced. Finally, the coefficient of the last term in the square brackets is adjusted in such a way that at the minimum

$$m_i \langle M^i_i \rangle = -\frac{2}{3} \langle S \rangle. \tag{49}$$

The latter expression is a consequence of the Konishi anomaly [7],

$$\frac{1}{8} D^2 \left( \bar{Q} e^V Q + \bar{\tilde{Q}} e^{-V} \tilde{Q} \right) = m Q \tilde{Q} + \frac{1}{16\pi^2} \text{Tr} W^2. \tag{50}$$

The superpotential in Eq.(45) has minima of two types. First, the chirally symmetric minimum at $S_* = 0$, observed in Ref. [19]. (All the quantities at the minima are marked by the asterisk.) The corresponding value of the superpotential is $\mathcal{W}_{*1} = 0$. Second, it has $N_c$ "standard" chirally asymmetric minima at

$$S_*^{N_c} = \det M_*^{\frac{N_c}{N_f - N_c}} = \left( -\frac{3}{2} \right)^{N_f} m_1 \ldots m_{N_f}. \tag{51}$$

The corresponding value of the superpotential is

$$\mathcal{W}_{*2} = -\frac{2N_c}{3} S_*. \tag{52}$$

Now, the Veneziano-Yankielowicz effective Lagrangian is nothing but a generalized Wess-Zumino model. The central charge is immediately obtained from the tree-level calculation, and is given by Eq.(17). Substituting Eq.(52) we find

$$\Sigma = 2 \left[ \mathcal{W}_{*2} - \mathcal{W}_{*1} \right] = -\frac{4N_c}{3} S_*. \tag{53}$$

The latter result is compatible with the general expression (43) at

$$C = \frac{N_f - N_c}{16\pi^2}, \tag{54}$$

provided one takes into account the Konishi anomaly (50). This value of constant $C$ is in agreement with Eq.(42).

Although, at first sight, it may seem that the derivation above relies on the presence of the additional, chirally symmetric vacuum at $S_* = 0$ (which was introduced in Ref. [19]; its existence is still to be confirmed), actually the calculation with minor changes stays intact if we consider the domain wall, that interpolates between two "standard" chirally asymmetric vacua. The final result for the constant $C$ is unchanged. It depends only on the general symmetry structure of the Veneziano-Yankielowicz effective Lagrangian.

Let us return to SU($N_c$) SQCD with $N_f$ flavors and ask the question why $N_f = N_c$ is special so that at $N_f < N_c$ the coefficient $C$ in front of the anomaly does not vanish, while at $N_f = N_c$ it does. At $N_f < N_c$ a superpotential for the (composite) field $M$ is generated non-perturbatively even in the absence of the tree-level superpotential $\mathcal{W}$ [25],

$$\mathcal{W}_{\text{n.p.}} \sim \left(\frac{\Lambda^{3N_c - N_f}}{\det M}\right)^{\frac{1}{N_c - N_f}}. \tag{55}$$

In our language this non-perturbatively generated superpotential is entirely due to the second, anomalous term in $\{Q_\alpha Q_\beta\}$. At $N_f = N_c$ the conserved $R$-charge argument forbids non-perturbative superpotential [25]. The vanishing of $C$ at $N_f = N_c$ is in one-to-one correspondence with the absence of any function of $\det M$ that would have appropriate $R$ charge.

It is clear, that two derivations of the central extension presented above are not independent. They both relate the anomaly in the anticommutator of supercharges to other anomalies of the theory. A direct computation of the anticommutator, say, by using the Bjorken-Johnson-Low limit [26], is more complicated due to complications with the gauge invariant supersymmetric regularization. Thus, the full diagrammatic calculation of the central extension is still missing. We have no doubts, however, that when it is carried out, our result (40) will be reproduced.

# III. BPS-SATURATED SOLUTIONS IN THE GENERALIZED WESS-ZUMINO MODEL

The generalized Wess-Zumino model is the theory of the type (6) with an extended set of the interacting chiral superfields $\Phi_i$, $i = 1, 2, ...$, and the superpotential $\mathcal{W}(\Phi_i)$, which will be assumed to be analytic function of the superfields and various parameters (coupling constants, mass terms and so on), if not stated to the contrary, see Eq. (56) below. Typically, one can think of $\mathcal{W}(\Phi_i)$ as of a polynomial. In certain instances of physical interest, however, one has to work with non-polynomial $\mathcal{W}(\Phi_i)$, see e.g. Ref. [22].

In the previous section we demonstrated that in many instances the extended classical field configurations lead to the existence of the central extension of $N = 1$ SUSY superalgebra (1).

The central extension turns out to be very important for the analysis of the theory in the sector with the domain walls or other topological defects, e.g. strings. As was already mentioned, the domain walls ruin the invariance of the theory with respect to the Poincaré group, and, with necessity, lead to (at least partial) supersymmetry breaking due to the algebra (15). The question whether or not supersymmetry is broken completely was shown [1] to be related to the appearance of the central extension. If the central charge vanishes the domain wall necessary completely breaks $N = 1$ supersymmetry. With a non-zero central charge two out of four supersymmetry generators may remain unbroken.

## A. Preservation of 1/2 of SUSY and BPS saturation

In Sect. II it was demonstrated that if the domain wall under consideration is BPS-saturated, the domain wall configuration preserves a part of supersymmetry. In the beginning of this section we will demonstrate that the inverse assertion is also true:

<u>*Theorem 2.*</u>   If a part of supersymmetry is preserved for the given domain wall, it is BPS-saturated, and its energy density $\varepsilon = |\Sigma|$.

Let us consider the generalized Wess-Zumino model, with a set of interacting chiral fields. Although we use a specific model, the results will be quite general, and, in particular, applicable to the gauge theories. The Lagrangian of the model has the form

$$\mathcal{L} = \sum_i \frac{1}{4} \int d^4\theta d^4x \Phi_i \bar{\Phi}_i + \left\{ \frac{1}{2} \int d^2\theta d^4x \mathcal{W}(\Phi_i) + H.C. \right\} . \tag{56}$$

We assume that the superpotential $\mathcal{W}$ is chosen in such a way, that the scalar potential has more than one minimum

$$\frac{\partial \mathcal{W}}{\partial \phi_i} = 0 \tag{57}$$

at

$$\{\phi_i\} = \{\phi_{i*}\}_1 \, , \, \{\phi_{i*}\}_2 \, , \ldots$$

where, as usual, the asterisk marks the values of the scalar fields $\phi_i$ at the minimum, and the subscript $1$, $2$, ... indicates that the set of minima includes several (isolated) points. To ease the notation, in what follows, we will omit the index $i$, referring to the scalar fields at the minima number $1$, $2$, ... as $\phi_{*1}$, $\phi_{*2}$ and so on.

The boundary conditions are

$$\phi \to \begin{cases} \phi_{*1} & \text{at } z \to -\infty \\ \phi_{*2} & \text{at } z \to +\infty \end{cases} . \tag{58}$$

The energy of the static field configuration with the given boundary condition (the domain wall) is

$$\mathcal{E} = A \int_{-\infty}^{+\infty} dz \left[ \sum_i \left( \partial_z \phi_i^+ \partial_z \phi_i \right) + \mathcal{V} \right] \tag{59}$$

where $\mathcal{V}$ is the scalar potential

$$\mathcal{V} = \sum_i \left| \frac{\partial \mathcal{W}}{\partial \phi_i} \right|^2 . \tag{60}$$

Since the energy is the sum of two non-negative terms, it can always be minimized. The field configuration with the boundary conditions (58) minimizing the energy $\mathcal{E}$ is the domain wall solution.

$$\partial_z^2 \phi_i = \frac{\partial}{\partial \phi_i^+} \left\{ \sum_j \left| \frac{\partial \mathcal{W}(\phi)}{\partial \phi_j} \right|^2 \right\}. \tag{61}$$

The domain wall solution, which we have chosen to be in the $xy$ plane, always exists since

$$\mathcal{V}(\phi_{*1}) = \mathcal{V}(\phi_{*1}) = 0. \tag{62}$$

It is convenient to denote the derivative with respect to $z$ as a dot over the corresponding letter. Moreover, we will arrange a set of $\phi$'s as a "vector",

$$\vec{\phi} = \begin{pmatrix} \phi_1 \\ \phi_2 \\ \cdots \end{pmatrix}. \tag{63}$$

Then Eq.(61) takes the form

$$\ddot{\vec{\phi}} = -\vec{\nabla}\mathcal{U} \tag{64}$$

where

$$\vec{\nabla} = \left\{ \frac{\partial}{\partial \phi_1^+}, \frac{\partial}{\partial \phi_2^+}, \cdots \right\} \tag{65}$$

the dot over letters denotes the derivative with respect to $z$ and

$$\mathcal{U} = -\sum_i \left| \frac{\partial \mathcal{W}}{\partial \phi_i} \right|^2. \tag{66}$$

Equation (64) is nothing but a complexified version of the Newtonian equation for the mechanical motion in the potential $\mathcal{U}$. For purely real solutions Eq.(64) reduces just to the regular Newton equation. The profile of $\mathcal{U}$ can be viewed as a mountain ridge with the summits at $\phi_{*1}, \phi_{*2}$ etc. The height of all summits is the same. The analog motion starts at one summit and ends at another.

So far the only new element introduced in the standard theory of the domain walls by supersymmetry is complexification. A more novel feature to which we proceed now is the existence of the special solutions satisfying the first order differential equations, rather then the second order equation (61).

In the spinorial notation the supersymmetry transformation has the form

$$\delta \bar{\psi}^{\dot{\alpha}} = \sqrt{2} \bar{\epsilon}^{\dot{\alpha}} F^+ - \sqrt{2} i (\partial_\mu \phi^+)(\bar{\sigma}^\mu)^{\dot{\alpha}\alpha} \epsilon_\alpha . \tag{67}$$

Assume that $\phi^+$ depends only on $z$ (i.e. $\phi^+$ is independent of $t$, $x$, $y$, the $(xy)$-plane oriented domain wall), and

$$\tau_1 \epsilon = i e^{i\alpha} \bar{\epsilon} \tag{68}$$

where $\epsilon$ and $\bar{\epsilon}$ are two-component columns

$$\epsilon = \begin{pmatrix} \epsilon^1 \\ \epsilon^2 \end{pmatrix}, \quad \bar{\epsilon} = \begin{pmatrix} \epsilon^{1*} \\ \epsilon^{2*} \end{pmatrix}, \tag{69}$$

and the factor $+i$ is singled out for convenience. (The matrix $\tau_1$ appears in Eq.(68) since both $\epsilon$'s in Eq. (68) have upper indices; therefore, in Eq. (67) we must substitute

$$\epsilon_\beta = -i(\tau_2)_{\beta\alpha} \epsilon^\alpha ,$$

see Appendix A, Eqs.(A11), (A12)). The solution of Eq.(68) is

$$\epsilon_0 = \begin{pmatrix} 1 \\ +i \end{pmatrix} e^{i \frac{\alpha}{2}} . \tag{70}$$

Then $\delta \bar{\psi}^{\dot{\alpha}} = 0$, provided that

$$\partial_z \phi^+ = -F^+ e^{-i\alpha} = \frac{\partial \mathcal{W}}{\partial \phi} e^{-i\alpha} . \tag{71}$$

If equation (71) is satisfied, two out of four supercharges will annihilate the domain wall, the corresponding part of supersymmetry will be preserved [1,2]. More exactly, Eq.(71) must hold for all fields $\phi_i$ from the set, with one and the same value of the phase $\alpha$.

Under the standard choice of the superpotential

$$\mathcal{W}_{*2} \equiv \mathcal{W}(\phi_{*2}) = \mathcal{W}\left[\phi_i^w(z = +\infty)\right] = \kappa \tag{72}$$

where $\kappa$ is real and positive (obviously $2\kappa = |\Sigma|$). This choice is convenient since under this choice we can fix the phase $\alpha$ in Eq.(71) at zero.

In the spirit of our mechanical analogy we can rewrite Eq.(71) in the following way

$$\dot{\vec{\phi}} = \vec{\nabla}\bar{\mathcal{W}}. \tag{73}$$

The "gradient" is defined in Eq.(65). This is nothing but a (complexified) equation of motion of the high-viscosity fluid whose inertia can be neglected. We will refer to Eqs.(73) as streamline or *creek* equations. The creek flow starts at one saddle point of $\bar{\mathcal{W}}$ (i.e. $\bar{\mathcal{W}}(\phi_{*1})$) and ends at another saddle point ($\bar{\mathcal{W}}(\phi_{*2})$).

It is quite clear that if Eq.(73) is satisfied the second-order relation (61) is satisfied too.* Indeed, start from Eq.(73) and differentiate both the left- and right-hand sides. We then get

$$\ddot{\phi}_i = \frac{\partial}{\partial z}\dot{\phi}_i = \frac{\partial}{\partial z}\left(\frac{\partial \bar{\mathcal{W}}}{\partial \phi_i^+}\right) = \sum_j \frac{\partial^2 \bar{\mathcal{W}}}{\partial \phi_i^+ \partial \phi_j^+}\dot{\phi}_j^+ = \sum_j \frac{\partial^2 \bar{\mathcal{W}}}{\partial \phi_i^+ \partial \phi_j^+}\frac{\partial \mathcal{W}}{\partial \phi_j} \tag{74}$$

which identically coincides with Eq.(61). The opposite is obviously not always true: not every solution of Eq.(61) is simultaneously the solution of the creek equation (71).

Now we are finally able to prove *Theorem 2* formulated above. Assume that we deal with a solution of the creek equations which then automatically preserves one half of SUSY. The energy of the field configuration is (see Eq.(59))

$$\mathcal{E} = 2A \int_{-\infty}^{+\infty} dz \left[\sum_i \dot{\phi}_i^+ \frac{\partial \bar{\mathcal{W}}}{\partial \phi_i^+}\right] = 2A\left(\bar{\mathcal{W}}_{*2} - \bar{\mathcal{W}}_{*1}\right) = A\left|\Sigma\right|. \tag{75}$$

The fact that $\mathcal{E}/A = |\Sigma|$ means, by definition, that the wall is BPS-saturated, q.e.d.

For completeness we give here an alternative proof (Refs. [15,16]) of *Theorem 1*, complementary to that given in Sect. II.B. Consider

$$\int_{-\infty}^{+\infty}\sum_i \left(\dot{\phi}_i - \frac{\partial \bar{\mathcal{W}}}{\partial \phi_i^+}\right)\left(\dot{\phi}_i^+ - \frac{\partial \mathcal{W}}{\partial \phi_i}\right) dz. \tag{76}$$

The integral (76) is obviously positive-definite for all $\phi_i$. At the same time, it can be rewritten as

$$\int_{-\infty}^{+\infty} \sum_i \left( \dot{\phi}_i^+ \dot{\phi}_i + \frac{\partial \bar{\mathcal{W}}}{\partial \phi_i^+} \frac{\partial \mathcal{W}}{\partial \phi_i} \right) - \int_{-\infty}^{+\infty} dz \frac{\partial}{\partial z} \left( \mathcal{W} + \bar{\mathcal{W}} \right) \geq 0. \tag{77}$$

---

*Analogous equations in non-supersymmetric models were considered in Ref. [14]

Minimizing the first term by varying $\phi_i$ we get

$$\varepsilon \geq |\Sigma|,$$

and the equality demands the domain wall function to satisfy

$$\dot{\phi}_i^+ = \frac{\partial \mathcal{W}}{\partial \phi_i}$$

Thus, this is necessary and sufficient condition.

### B. Properties of the BPS-saturated walls

Within our convention regarding the superpotential we can prove a stronger statement.

*Theorem 3.* Let the BPS-saturated domain wall exist. Then $\mathcal{W}[\phi_i^w(z)]$ is a real positive number for all $z$, provided our standard convention on the superpotential is adopted.

This relation is a direct consequence of Eq.(71). Indeed, let us multiply both sides by $\dot{\phi}$ and integrate over $z$ in the interval from $-\infty$ to the current value of $z$. Then we get

$$\{\mathcal{W}[\phi_i^w(z)] - \mathcal{W}_{*1}\} = \int_{-\infty}^{z} dz |\dot{\phi}|^2 \geq 0. \tag{78}$$

This proves the theorem, since according to our standard choice of the superpotential $\mathcal{W}_{*1} = 0$. A particular case of this result was used previously [22].

The superfield describing the domain wall so far has only bosonic components,

$$\Phi_w = \phi_w + \theta^2 F_w = \phi_w - \theta^2 \frac{\partial \mathcal{W}(\phi)}{\partial \phi} \Big|_{\phi = \phi_w}. \tag{79}$$

If we add the fermion zero modes, generated by those supertransformations that act on the wall non-trivially, we will get the wall superfield, which has all three components, the lowest, the middle, and the $F$ component. In the case of the BPS saturation the wall superfield satisfies the full supersymmetric equation of motion,

$$\frac{1}{4}D^2\Phi_i = \frac{\partial \bar{\mathcal{W}}(\bar{\Phi})}{\partial \bar{\Phi}_i};  \qquad (80)$$

for further details see Sect. IV.A.

It is obvious that if the central charge $\Sigma = 0$ the corresponding domain wall breaks supersymmetry completely [1]. From *Theorem 1* it is clear that the BPS-saturated solutions preserving a part of SUSY can exist only provided $\Sigma \neq 0$. Unfortunately, the non-vanishing central charge is not sufficient to ensure the existence of the BPS-saturated domain walls.

Let us examine the example suggested in Ref. [1]: the generalized Wess-Zumino model with two chiral superfields, $X$ and $\Phi$, and the superpotential

$$\mathcal{W} = X\left(\mu^2 - \lambda\Phi^2\right) \qquad (81)$$

(cf. Eq.(7)). This superpotential has two minima:

$$\left\{\phi = \frac{\mu}{\sqrt{\lambda}}, \chi = 0\right\} \text{ and } \left\{\phi = -\frac{\mu}{\sqrt{\lambda}}, \chi = 0\right\} \qquad (82)$$

where $\chi$ is the lowest component of the superfield $X$. The corresponding values of the superpotential at the minima vanish. Hence, the central charge $\Sigma = 0$ in the model (81), and the creek equations have no solution. At the same time, the classical equations of motion (61) for the field $\phi$ identically coincide with that of the model (7) provided we put $X = 0$. The solution is of course the same, as in the model (7), with the energy density

$$\varepsilon = \frac{8\mu^3}{3\sqrt{\lambda}}. \qquad (83)$$

(See Sect. 4 for further details; it is assumed for the time being that $\mu$ and $\lambda$ are real positive parameters.) As we know, the solution of Eq.(61) always exists. The fact that $\varepsilon > 0$ while $\Sigma = 0$ shows that the domain wall is not BPS-saturated, and, correspondingly, SUSY is completely broken.

Now, let us deform the model in such a way as to generate a non-vanishing central charge,

$$\mathcal{W}(\alpha) = X\left(\mu^2 - \lambda\Phi^2\right) + \frac{\alpha}{3}\Phi^3. \qquad (84)$$

The values of the scalar fields at the minima become

$$\left\{\phi = \frac{\mu}{\sqrt{\lambda}},\ \chi = \frac{\alpha}{2\lambda}\frac{\mu}{\sqrt{\lambda}}\right\}$$

and

$$\left\{\phi = -\frac{\mu}{\sqrt{\lambda}},\ \chi = -\frac{\alpha}{2\lambda}\frac{\mu}{\sqrt{\lambda}}\right\}.$$

At the minima

$$\mathcal{W}(\alpha)_{*1} = -\frac{\alpha}{3}\frac{\mu^3}{\lambda\sqrt{\lambda}},\quad \mathcal{W}(\alpha)_{*2} = +\frac{\alpha}{3}\frac{\mu^3}{\lambda\sqrt{\lambda}} \tag{85}$$

The central charge

$$\Sigma(\alpha) = 2\Delta\mathcal{W}(\alpha)_* = \frac{4\alpha}{3}\frac{\mu^3}{\lambda\sqrt{\lambda}} \neq 0, \tag{86}$$

(the parameter $\alpha$ is assumed to be real and positive ). The domain wall solution of the model (81) will also be deformed. It is clear, however, that as long as $\alpha/\lambda \ll 1$ the deformation is small, and the energy density $\varepsilon(\alpha)$ is close to Eq.(83). Thus, for sufficiently small $\alpha/\lambda$

$$\varepsilon(\alpha) \gg |\Sigma(\alpha)|, \tag{87}$$

and the domain wall can not be BPS-saturated.

As a matter of fact, it is not difficult to prove a stronger statement. Assume that at some positive value of $\alpha$ the domain wall in the model at hand is BPS-saturated. Then it satisfies the creek equations

$$\dot{\chi}^+ = \mu^2 - \lambda\phi^2,\quad \dot{\phi}^+ = \alpha\phi^2 - 2\lambda\chi\phi. \tag{88}$$

Multiply the first equation by $\dot{\phi}$ and integrate over $z$ from $-\infty$ to $+\infty$,

$$I_1 = \int_{-\infty}^{+\infty} dz\,\dot{\phi}\dot{\chi}^+ = \int_{-\infty}^{+\infty} dz\,\frac{\partial}{\partial z}\left(\mu^2\phi - \frac{\lambda}{3}\phi^3\right) = \frac{4}{3}\frac{\mu^3}{\sqrt{\lambda}}. \tag{89}$$

The integral

$$I_2 = \int_{-\infty}^{+\infty} dz\,\left(\dot{\phi} - \dot{\chi}\right)\left(\dot{\phi}^+ - \dot{\chi}^+\right) =$$
$$\int_{-\infty}^{+\infty} \left(\dot{\phi}\dot{\phi}^+ + \dot{\chi}\dot{\chi}^+\right) dz - 2I_1 \tag{90}$$

is obviously non-negative. The first integral on the right-hand side

$$\int_{-\infty}^{+\infty} \left( \dot\phi\dot\phi^+ + \dot\chi\dot\chi^+ \right) dz = \frac{1}{2} \int_{-\infty}^{+\infty} \left( \dot\phi\dot\phi^+ + \dot\chi\dot\chi^+ + \mathcal{V} \right) dz = \frac{\varepsilon}{2}. \tag{91}$$

If the domain wall is PBS-saturated then

$$\varepsilon = |\Sigma|. \tag{92}$$

Hence in this case we must have

$$\frac{1}{2}|\Sigma| \geq 2I_1 \tag{93}$$

or

$$\frac{2\alpha}{3} \frac{\mu^3}{\lambda\sqrt{\lambda}} \geq \frac{8}{3} \frac{\mu^3}{\sqrt{\lambda}}. \tag{94}$$

We conclude that the domain wall at hand can *not* be BPS- saturated at $\alpha/\lambda < 4$. If there is any chance to make it BPS-saturated $\alpha/\lambda$ must be $\geq 4$.

Actually, there is no chance. The theory under consideration has two fermion fields, $\psi_\Phi$ and $\psi_X$. Consider the fermion mass matrix $\partial^2 \mathcal{W}/\partial\Phi_i\partial\Phi_j$ in either of two vacua (85). After diagonalization of the mass matrix we find two independent diagonal linear combinations of $\psi_\Phi$ and $\psi_X$ (they are the same in both vacua). The sign of the mass term *changes* in passing from one vacuum to another. According to the index theorem [27], the Dirac equation for each of the diagonal combination will have a zero mode. At the same time only one linear combination of the bosonic fields has a zero mode. Hence, SUSY is broken completely on the domain wall for any $\alpha$. The absence of the BPS-saturated solutions for any $\alpha$ for real values of the fields is readily seen from the consideration of the profile of $\mathcal{W}$.

*Conjecture.* If in the generalized Wess-Zumino model with the polynomial potential the domain wall is BPS-saturated at some values of parameters in the superpotential, it continues to be BPS-saturated for all non-singular deformations of parameters as long as the central charge remains finite (neither infinity, nor zero), and

$$\det\left\{ \frac{\partial^2 \mathcal{W}}{\partial\Phi_i\partial\Phi_j} \right\}_{\Phi=\Phi_*}$$

does not cross zero. If at some values of the parameters the domain wall is not BPS-saturated, it can not become BPS-saturated under arbitrary non-singular deformations of the parameters. (Adding to the superpotential terms of higher powers in the fields which were originally absent is a singular deformation). The latter statement is valid provided there is no level crossing, i.e. the energy densities of two domain walls, $\varepsilon_1(\lambda)$ and $\varepsilon_2(\lambda)$, do not collide at some $\lambda$ (here $\lambda$ is a generic notation for the set of parameters in the superpotential).

The simplest example of the BPS-saturated domain wall is provided by the minimal Wess-Zumino model, see [1] and Sect. 4.

Unfortunately, we were unable to find a sufficient condition for the existence of the BPS-saturated domain walls in the general case. In Ref. [22] the BPS-saturated walls were found in the Veneziano-Yankielowicz Lagrangian which gives an effective description of the supersymmetric Yang- Mills theories (with or without matter).

If the central charge is non-vanishing, the parameters of the superpotential are *purely real*, and one decides to limit one's searches for the BPS-saturated domain walls by purely real solutions, one can take advantage of the fact that Eq.(71) is analogous to the one describing the motion of the high-viscosity liquid on the multidimensional profile given by the superpotential $\mathcal{W}$. This mechanical analogy and our intuition in solving mechanical problems, helps in establishing the condition for the existence of the BPS-saturated solutions in this case. Making use of the mechanical analogy, we can infer that if the superpotential has a set of extrema, and at least one of them is either maximum or minimum, then there always exists a BPS-saturated domain wall solution, and 1/2 of SUSY is preserved. (If all extremum points are saddle points, the BPS-saturated solution may not exist.)

It is instructive to present here the corresponding example – a model in which the BPS-saturated wall obviously exists for real values of the parameters in the superpotential, and it remains to be BPS-saturated when one analytically continues the parameters of the superpotential in the complex plane provided that the central charge stays finite (does not go to zero or infinity). The model we keep in mind has the following superpotential

$$\mathcal{W}(\Phi, X) = -\alpha X + \frac{\beta}{2}\Phi^2 + \frac{\gamma}{3}(X - \Phi)^3, \tag{95}$$

where $\alpha, \beta$ and $\gamma$ are parameters. To begin with, we choose all three parameters real and positive. Then the solution of the creek equations can be sought for in the class of the real functions. By inspecting the profile of $\mathcal{W}$ it immediately becomes clear, that this profile has one maximum and one saddle point, and the creek equations do have a solution. The analytic form of the solution is unknown, but it exists, for sure. Moreover, the extrema of $\mathcal{W}$ are

$$\phi_{*1} = \frac{\alpha}{\beta}, \quad \chi_{*1} = \frac{\alpha}{\beta} + \sqrt{\frac{\alpha}{\gamma}}$$

(maximum) and

$$\phi_{*2} = \frac{\alpha}{\beta}, \quad \chi_{*2} = \frac{\alpha}{\beta} - \sqrt{\frac{\alpha}{\gamma}}$$

(the saddle point). The central charge

$$\Sigma = 2\Delta\mathcal{W}_* = \frac{8}{3}\alpha\sqrt{\frac{\alpha}{\gamma}} \tag{96}$$

is non-singular and non-zero provided $\alpha$ and $\gamma \neq 0, \infty$. The determinant of the matrix of the second derivatives of the superpotential at the extrema is

$$2\beta\gamma\sqrt{\frac{\alpha}{\gamma}}, \quad \text{and} \quad -2\beta\gamma\sqrt{\frac{\alpha}{\gamma}},$$

respectively. We can now move the parameters into the complex plane keeping $\beta \neq 0$; the BPS-saturated wall will persist. The wall energy density will be determined by the absolute value of (96).

In summary, the existence of the central extension with a non-vanishing central charge is a necessary condition for domain wall to be BPS-saturated [1]. The absence of the central extension (i.e. vanishing central charge) necessarily leads to the complete breaking of supersymmetry, while in the presence of central extension supersymmetry may or may not be broken completely.

**C. Three or more vacua**

So far we essentially disregarded the possibility that one and the same model can have more than two vacuum states, and, correspondingly, a variety of distinct domain walls.

Let us start, however, with a remark referring to models with one pair of supersymmetric vacua. One and the same model can have simultaneously two types of the domain walls interpolating between the same vacua. The picture becomes transparent if one limits oneself to superpotentials with real parameters and real solutions of the creek equations.

The first type of domain walls appears as the solution of the classical (second-order) equations of motion (61) in the bosonic sector. These solutions do not satisfy the creek equations. It is clear, that supersymmetry is not preserved on such walls. Solutions of the second type, are those of the first order differential equations (80), and they are partially supersymmetric by construction. Classically, both types of solutions may coexist peacefully, and be stable. The energy density in the first class is higher than in the second, and they become unstable quantum-mechanically. The "false" domain wall will transform into the BPS-saturated through the formation of "bubbles" on the wall which will grow in an explosive manner.

If there are more than two vacua, the picture may become even more versatile. Some of the vacua may be connected by BPS-saturated walls, others by non-saturated walls. Let us assume for simplicity that the model at hand has three vacuum states, and all are connected by the BPS-saturated walls. We would like to consider a superposition of two walls – a field configuration that first interpolates between the first and the second vacuum, and then interpolates between the second and the third. If these two walls are separated by a finite distance in the $z$ direction, the corresponding field configuration is not the exact solution (it only approaches it when the separation tends to infinity). The behavior of the superimposed walls crucially depends on the corresponding central charges (which are generically complex). Let us denote by $\Sigma_{ij}$ the central charge for the wall configuration interpolating between the $i$-th and $j$-th vacua. If all $\Sigma_{ij}$ lie on one line in the complex plane and are ordered along this line then, obviously, the superposition of walls 12 and 23 has higher energy than the BPS-saturated wall 13 would have. However, the infinitely distant walls 12 and 23 will have exactly the same energy as the BPS-saturated wall 13. This means, that there is a repulsion between walls 12 and 23 which will eventually push them out at infinite separation. If there is an independent BPS-saturated interpolation between 1 and 3, these two types of configurations will coexist.

The above case is clearly degenerate. If the central charges do not lie on one line and form a triangular, with the corresponding inequality,

$$|\Sigma_{12}| + |\Sigma_{23}| \geq |\Sigma_{13}|$$

the infinitely distant walls 12 and 23 will have larger energy as the BPS-saturated wall 13. This will lead to the wall attraction at large distances. The walls 12 and 23 will collapse, forming a lower $\varepsilon$ state, the wall 13.

It is interesting that many elements of this rich picture are actually realized in supersymmetric $SU(N)$ Yang-Mills theory [22]. In this model there exist $N$ domain walls connecting $N$ chirally asymmetric vacua with the chirally symmetric one [19] at $\langle\lambda\lambda\rangle = 0$. Let us call them *primary*. These walls are trivially BPS-saturated. In the $SU(2)$ model there are three vacua, and all central charges are real. Hence, if we superimpose two primary walls, we will get a repulsion at large distances. For $SU(3)$ and higher groups $SU(N)$ there are also $N(N-1)/2$ walls connecting directly various chirally asymmetric vacua, which also seem to be BPS-saturated [28]. The primary walls experience strong attraction at large distances; the walls connecting the chirally asymmetric vacua can be viewed as bound states of the primary walls.

### D. BPS-saturated strings

The very existence of the walls is due to the fact that our space is not compact – the dynamical fields at all infinitely distant points need not have one and the same value. We can use the very same idea to consider other topological defects, in the spirit of what was done in Ref. [1]. Our consideration here is conceptually close to that given in Sects. 3 and 5 of Ref. [1], but we will focus on those defects which are BPS-saturated, i.e. 1/2 of the original SUSY is preserved in the corresponding background field. The configurations to be discussed are string-like solutions, static and $z$ independent. They spontaneously break two out of four Lorentz translations, so that the resulting dynamical system has $D = 1+1$.

Let us orient the string along the $z$ direction, so that all bosonic fields in the solution we are going to build will depend only on $x$ and $y$. Introduce

$$\zeta = x + iy \text{ and } \bar{\zeta} = x - iy.$$

The central extension of the superalgebra, see Eq. (13), now takes the form

$$\{Q_\alpha\, Q_\beta\} = 4L\left\{\int d\bar{\zeta}\,\frac{1}{2}(1-\tau_3)_{\alpha\beta}\left[\bar{\mathcal{W}}[\bar{\Phi}_k(\zeta,\bar{\zeta})] - \frac{1}{3}\sum_i \bar{\Phi}_i\frac{\partial \bar{\mathcal{W}}[\bar{\Phi}_k(\zeta,\bar{\zeta})]}{\partial \bar{\Phi}_i}\right] - \right.$$

$$\left.\int d\zeta\,\frac{1}{2}(1+\tau_3)_{\alpha\beta}\left[\bar{\mathcal{W}}[\bar{\Phi}_k(\zeta,\bar{\zeta})] - \frac{1}{3}\sum_i \bar{\Phi}_i\frac{\partial \bar{\mathcal{W}}[\bar{\Phi}_k(\zeta,\bar{\zeta})]}{\partial \bar{\Phi}_i}\right]\right\}_{\bar{\theta}=0}. \tag{97}$$

where $L$ is the length of the string, and the integrals run over the large circle in the $\zeta$ plane. The expression in the braces is the central charge. This expression assumes that the fields $\Phi_i$ are non-singular at finite distances, and the superpotential $\mathcal{W}$ is non-singular at finite values of the fields. Introduction of the gauge fields results in obvious modifications, with the anomaly term appearing on the right-hand side.

What supercharges are conserved on the BPS-saturated strings? To answer the question we must go back to the general supertransformation law (67). If the parameter of the supertransformation is chosen as follows

$$\epsilon \equiv \begin{pmatrix} \epsilon^1 \\ \epsilon^2 \end{pmatrix} = \begin{pmatrix} 0 \\ \exp(i\pi/4) \end{pmatrix}, \tag{98}$$

then $\delta\bar{\psi} = 0$ provided

$$\partial \phi_i = \frac{1}{2}\frac{\partial \bar{\mathcal{W}}}{\partial \phi_i^+}. \tag{99}$$

Here
$$\partial \equiv \partial/\partial\zeta = \frac{1}{2}\left(\frac{\partial}{\partial x} - i\frac{\partial}{\partial y}\right)$$
and
$$\bar{\partial} \equiv \partial/\partial\bar{\zeta} = \frac{1}{2}\left(\frac{\partial}{\partial x} + i\frac{\partial}{\partial y}\right).$$

We could have introduced an arbitrary phase in Eq. (99), much in the same way as in the discussion following Eq. (67). This phase can always be "unwind" by an appropriate phase rotation of the superpotential. Needless to say that interchanging the lower and upper components of $\epsilon$ we would arrive at the first-order equations defining the "anti-string".

We will continue to call (99) the creek equations, although the "time" variable is now complex.

It is seen that the creek equations (99) corresponding to BPS saturation in the string case, are modified in a minimal way compared to the domain wall problem. As in the latter, there are several alternative lines of reasoning allowing one to obtain the creek equations. Say, we could have started from the analog of Eq. (76) and consider

$$\tilde{I} = 4\int \sum_i \left(\partial\phi_i - \frac{1}{2}\frac{\partial\bar{\mathcal{W}}}{\partial\phi_i^+}\right)\left(\bar{\partial}\phi_i^+ - \frac{1}{2}\frac{\partial\mathcal{W}}{\partial\phi_i}\right) d^2\zeta. \tag{100}$$

This integral is obviously positive-definite; it becomes zero only if each of the brackets vanishes. This is the condition of the BPS saturation. If Eq. (99) holds, then the standard equations of motion obviously hold too. Indeed,

$$\left(\frac{\partial^2}{\partial x^2} + \frac{\partial^2}{\partial y^2}\right)\phi_i = 4\bar{\partial}\partial\phi_i = 2\bar{\partial}\frac{\partial\bar{\mathcal{W}}}{\partial\phi_i^+} = \frac{\partial^2\bar{\mathcal{W}}}{\partial\phi_i^+\partial\phi_j^+}\frac{\partial\mathcal{W}}{\partial\phi_j} = \frac{\partial\mathcal{V}}{\partial\phi_i^+}. \tag{101}$$

Up to a total derivative $\tilde{I}$ reduces to the string tension $\sigma$ minus the central charge. If the latter is non-zero, the first-order equations (99) may have a solution. Then the string tension will be exactly equal to the central charge.

The statement above requires an immediate reservation: $\tilde{I}$ reduces to the string tension minus the central charge provided we discard total derivatives. Presumably, this can be achieved by adding certain total derivatives directly in the Lagrangian. The issue is not completely clear at the moment, further investigation is needed.

Let us consider a particular example, the model with the superpotential

$$\mathcal{W} = X - \frac{1}{\sqrt{2}} X^2 \Phi. \tag{102}$$

For simplicity the mass parameter in front of $X$ is put to unity. In the conventional sector, without topological defects, this model has a runaway vacuum,

$$\mathcal{V} = \frac{1}{2}|X^2|^2 + |1 - \sqrt{2} X\Phi|^2,$$

corresponding to $X \to 0$, $\Phi \to \infty$, much in the same way as in $SU(2)$ SQCD with one flavor and the vanishing matter mass term. It is easy to see that the creek equations (99) have the solution

$$X = \zeta(1+\zeta\bar\zeta)^{-1}, \quad \Phi = \frac{1}{\sqrt{2}}\bar\zeta\left[1 + (1+\zeta\bar\zeta)^{-1}\right]. \tag{103}$$

This solution has a constant volume energy density due to the kinetic term of the $\Phi$ field, as in Sect. 5 of Ref. [1]. This depends on the definition (i.e. on whether or not we add total derivatives). Discarding this constant volume energy density we observe a typical string-like distribution of energy, with a finite string tension. The central charge is non-vanishing since at large $|\zeta|$ the linear part of the superpotential $\mathcal{W}$ behaves as $1/\bar\zeta$. (The cubic part can be ignored as is evident from Eq. (97)).

## IV. MINIMAL WESS-ZUMINO MODEL

In this section we carry out the construction of supersymmetric theory in $(2+1)$ dimensions, obtained by dimensional reduction of the minimal Wess-Zumino model in $(3+1)$ dimensions on the domain wall solution. We construct an explicitly supersymmetric Lagrangian with the infinite set of interacting superfields. Each superfield realizes an irreducible representation of $N = 1$ supersymmetry in $(2+1)$ dimensions. We, then, compute the quantum corrections to the shape of the domain wall. In Sect. 5 we will use this information to obtain the first subleading correction in the asymptotic behavior of the multiparticle production amplitudes at threshold when the number of particles tends to infinity.

## A. Classical solution

The Lagrangian for the minimal Wess-Zumino model was written in Eq. (6). For our present purposes it is more convenient to define the parameters in the superpotential as follows:

$$\mathcal{W}(\Phi) = \frac{m^2}{\lambda}\Phi - \frac{\lambda}{3}\Phi^3, \tag{104}$$

where

$$\Phi = \phi + \sqrt{2}\theta\psi + \theta\theta F \tag{105}$$

is the only chiral superfield of the minimal model. It consists of one complex scalar field $\phi$, one chiral fermion $\psi$, and an auxiliary field $F$.

Although at first sight it might seem that the choice of the superpotential is not general, actually it is not difficult to see that any *renormalizable* model with one chiral superfield can be reduced to Eq. (104) by exploiting the freedom we have in defining the superfield. After $\Phi$ is appropriately shifted and rotated we arrive to Eq. (104), with the real values of parameters, no matter which cubic superpotential we started from.

Lagrangian (6) has a discrete $Z_2$ symmetry group (cf. Eq.(9)), which is spontaneously broken, so that there exists a classical solution $\Phi_w$ of equations of motion:

$$\frac{1}{4}\bar{D}^2\bar{\Phi}_w = \frac{\partial\mathcal{W}(\Phi)}{\partial\Phi}\Big|_{\Phi=\Phi_w} \tag{106}$$

where

$$\Phi_w = \phi_w + \theta\theta F_w = \phi_w - \theta\theta\bar{\mathcal{W}}'(\phi_w^+), \tag{107}$$

and $F = -\bar{\mathcal{W}}' = -\partial\bar{\mathcal{W}}/\partial\phi^+$.

Here $\phi_w$ is the real field $\phi_w = \phi_w^+$

$$\phi_w = \frac{m}{\lambda}\tanh(mz) \tag{108}$$

satisfying

$$\frac{\partial \phi_w(z)}{\partial z} = \mathcal{W}'(\phi_w). \tag{109}$$

The domain wall was chosen to lie in the $xy$ plane.

Note that in the minimal Wess-Zumino model (when the solution is sought for in the class of real functions $\phi(z)$) it is always possible to reduce the second order differential equations of motion (61) to the creek equations, by exploiting the integral of motion (conservation of energy). Thus, the fact that the domain wall in the minimal Wess-Zumino model preserves 1/2 of SUSY is rather trivial. The domain wall is, then, BPS-saturated and its energy density coincides with the topological charge

$$\varepsilon = |\Sigma| = \frac{8m^3}{3\lambda^2}. \tag{110}$$

Let us discuss now the "other" half of supersymmetry, namely those supergenerators that are explicitly broken by the wall. The domain wall at hand satisfies the creek equation (109) (the wall profile is purely real). The generic supersymmetry transformation (67) in this case can be identically rewritten as

$$\delta\bar{\psi} = -\sqrt{2}\left(\frac{\partial \phi_w^+}{\partial z}\right)(\bar{\epsilon} + i\tau_1 \epsilon), \tag{111}$$

where in the remainder of this subsection all spinors are assumed to be taken with the upper indices (which will not be shown explicitly).

Applying the broken symmetry generators to the domain wall superfield $\phi_w$ produces zero modes. For instance, action of the broken generator of translation along the $z$ axis gives the bosonic zero mode. We are interested now in the fermion zero mode. The super-transformation generating the fermion zero mode corresponds to the parameter

$$\epsilon = \begin{pmatrix} 1 \\ -i \end{pmatrix}, \tag{112}$$

so that $\tau_1 \epsilon = -i\bar{\epsilon}$, and

$$\psi_{\text{z.m.}} = -2\sqrt{2}\bar{\epsilon}\left(\frac{\partial \phi_w^+}{\partial z}\right). \tag{113}$$

To describe simultaneously all zero modes, fermionic and bosonic, it is convenient to introduce the wall superfield $\Phi_w$ much in the same way as it was done for the instanton superfield in Ref. [29] (see especially Sect. 3 in this paper). To this end we first need to introduce the collective coordinates. The bosonic collective coordinate is $z_0$, the position of the center of the wall. In the consideration above it was assumed that the wall center is at the origin. Clearly, this is not the generic case; the wall center can be at any point on the $z$ axis. To get the wall superfield it is necessary to restore $z_0$. Now, we additionally introduce the fermion collective coordinate $\bar\theta_0$, with the transformation property

$$\delta\bar\theta_0 = \bar\epsilon + i\tau_1\epsilon. \tag{114}$$

Although this collective coordinate is formally written as a two-component spinor, actually it represents *one complex* Grassmann number,

$$\bar\theta_0 = \begin{pmatrix} \eta \\ i\bar\eta \end{pmatrix}, \tag{115}$$

and $\delta\eta = \bar\epsilon^1 + i\epsilon^2$. Needless to say that there exists also a conjugated collective coordinate, $\theta_0$.

The transformation properties of $\bar\theta_0$ are rather peculiar. Actually, for building the wall superfield $\bar\Phi_w$ we will only need to consider the differences $z_R - z_0$ and $\bar\theta - \bar\theta_0$. (Naturally, in the conjugated superfield $\Phi_w$ one deals with $z_L - z_0$ and $\theta - \theta_0$.)

Then,

$$\delta(\bar\theta - \bar\theta_0) = -i\tau_1\epsilon, \tag{116}$$

and

$$\delta(z_R - z_0) = 2i[\epsilon\tau_3(\bar\theta - \bar\theta_0)]. \tag{117}$$

It is easy to check that the invariant interval has the form

$$(\Delta z)_{\text{inv}} = (z_R - z_0) - i(\bar\theta - \bar\theta_0)\tau_2(\bar\theta - \bar\theta_0). \tag{118}$$

Now, if we take the domain wall solution $\phi_w^+$ of the pre-supersymmetry era, given in Eq. (108), and substitute there $z$ by $(\Delta z)_{\text{inv}}$, we will get the full wall superfield $\bar{\Phi}_w$,

$$\bar{\Phi}_w(z_R, \bar{\theta}) = \frac{m}{\lambda}\tanh[m(\Delta z)_{\text{inv}}], \tag{119}$$

which contains the classical solution itself, and the zero modes. The fermion zero mode emerges when one expands $(\Delta z)_{\text{inv}}$ appearing in tanh above, keeping the term linear in $\bar{\theta}$. The term quadratic in $\bar{\theta}$ will automatically produce the $F^+$ component. The $\bar{\theta}_0^2$ part of $(\Delta z)_{\text{inv}}$ corresponds to shifting the domain wall center. The expression for $\Phi_w$ is similar.

The meaning of the parameters $\bar{\theta}_0, \theta_0$ (fermionic collective coordinates) is now clear.

### B. Quantum fluctuations

To obtain equations that govern the quantum fluctuations around the classical solution we rewrite Lagrangian (6) in components and expand it in "quantum" parts of the fields around the classical configuration, keeping only the quadratic in the "quantum" fields terms,

$$\mathcal{L} = \int d^4x \Big\{ (\partial_\mu a)^2 + (\partial_\mu b)^2 - (a^2 + b^2)[\mathcal{W}''(\phi_w)]^2 - (a^2 - b^2)\mathcal{W}'(\phi_w)\mathcal{W}'''(\phi_w)$$
$$+ \psi^\alpha i\partial_{\alpha\dot\alpha}\bar{\psi}^{\dot\alpha} - \frac{1}{2}\mathcal{W}''(\phi_w)\left(\bar{\psi}\bar{\psi} + \psi\psi\right)\Big\}, \tag{120}$$

here

$$a = \text{Re}(\phi - \phi_w); \quad b = \text{Im}(\phi - \phi_w).$$

Potential terms are different for real and imaginary parts of the scalar field, for the classical background is real. This gives the following set of equations of motion:

$$\{\partial'^2 + \mathcal{O}^2\mathcal{O}^1(z)\}a = 0,$$
$$\{\partial'^2 + \mathcal{O}^1\mathcal{O}^2(z)\}b = 0,$$
$$i\hat{\partial}\Psi^+ + \mathcal{O}^2(z)\Psi^- = 0,$$
$$i\hat{\partial}\Psi^- + \mathcal{O}^1(z)\Psi^+ = 0 \tag{121}$$

where we introduced a short-hand notation

$$\mathcal{O}^2(z) = \{\partial_z + \mathcal{W}''(\phi_w)\},$$

$$\mathcal{O}^1(z) = \{-\partial_z + \mathcal{W}''(\phi_w)\},$$

$$\Psi_\alpha^+ = \left[\frac{\psi_\alpha + i(\sigma^3)_{\alpha\dot\alpha}\bar\psi^{\dot\alpha}}{\sqrt{2}}\right], \quad \Psi_\alpha^- = \left[\frac{\psi_\alpha - i(\sigma^3)_{\alpha\dot\alpha}\bar\psi^{\dot\alpha}}{i\sqrt{2}}\right],$$

$$\hat\partial_\alpha^\beta = (\partial^0\tau^3 - i\partial^1\tau^2 + i\partial^2\tau^1)_{\alpha\beta},$$

$$\partial' = \{\partial_0, \partial_1, \partial_2\} \tag{122}$$

where $(\tau^i)^{\alpha\beta}$ are the Pauli matrices.

Since the classical solution $\phi_w$ depends only upon coordinate $z$, the $z$ dependence of the fields can be factorized

$$\begin{aligned}
a(x) &= \sum_n \sqrt{\lambda_n}\, a_n(z)\, a_n(x^1, x^2, x^0),\\
\Psi_n^+(x) &= \sum_n \sqrt{\lambda_n}\, a_n(z)\, \Psi_n^+(x^1, x^2, x^0),\\
b(x) &= \sum_n \sqrt{\kappa_n}\, b_n(z)\, b_n(x^1, x^2, x^0),\\
\Psi_n^-(x) &= \sum_n \sqrt{\kappa_n}\, b_n(z)\, \Psi_n^-(x^1, x^2, x^0)
\end{aligned} \tag{123}$$

where the factors $\sqrt{\lambda_n}$ or $\sqrt{\kappa_n}$ are inserted for proper normalization ( $\lambda_n$ and $\kappa_n$ are the eigenvalues, see Eqs.(124), (127) below).

Functions $a_n(z)$, $b_n(z)$ satisfy the following equations:

$$\begin{aligned}
\mathcal{O}^1(z)a_n(z) &= \kappa_n b_n(z),\\
\mathcal{O}^2(z)b_n(z) &= \lambda_n a_n(z).
\end{aligned} \tag{124}$$

They are eigenfunctions of two different operators

$$\begin{aligned}
\mathcal{O}^2\mathcal{O}^1(z)a_n(z) &= m_n^2 a_n(z),\\
\mathcal{O}^1\mathcal{O}^2(z)b_n(z) &= m_n^2 b_n(z),
\end{aligned} \tag{125}$$

where $m_n^2 = \lambda_n\kappa_n$, and, thus, can not be normalized to unity simultaneously. Making use of Eq.(124) we easily obtain (brackets stay for the integration over $z$)

$$\kappa_n \langle b_n\, b_n \rangle = \lambda_n \langle a_n\, a_n \rangle, \tag{126}$$

and the normalization condition is

$$\langle b_n b_n \rangle = \frac{1}{\kappa_n}, \quad \langle a_n a_n \rangle = \frac{1}{\lambda_n}. \tag{127}$$

We substitute the expansion (123) into the Lagrangian (120), integrate over $z$ to obtain three-dimensional Lagrangian for the fluctuations on the surface of the domain wall. Each given mode of Eq.(124) will correspond to a particle with the mass given by the corresponding eigenvalue. Since we carried out the expansion around the classical solution only up to the second order in fields we obtain a free theory of two bosonic and four fermionic components,

$$\mathcal{L} = \int d^3 x \sum_n \left\{ (\partial_\mu a_n)^2 + (\partial_\mu b_n)^2 - m_n^2(a_n^2 + b_n^2) \right. \\ \left. + \frac{1}{2}\alpha_n(i\hat{\partial} - m_n)\alpha_n + \frac{1}{2}\beta_n(i\hat{\partial} + m_n)\beta_n \right\} \tag{128}$$

where

$$m_n = \sqrt{\lambda_a^n \lambda_b^n}; \quad \alpha_n = \frac{\Psi_n^+ + i\Psi_n^-}{\sqrt{2}}; \quad \beta_n = \frac{\Psi_n^+ - i\Psi_n^-}{\sqrt{2}}, \tag{129}$$

and $\hat{\partial}$ is given in Eq.(122). The summation convention used throughout the section is

$$\alpha\alpha = \alpha^\beta \alpha_\beta; \quad \alpha\hat{\partial}\alpha = \alpha^\beta \hat{\partial}_\beta^\gamma \alpha_\gamma.$$

The Lagrangian (128) is obviously supersymmetric. The pairs $(\tilde{a}, \alpha)$ and $(\tilde{b}, \beta)$, where the tilded fields are some orthogonal linear combinations of the fields $a$ and $b$, form two irreducible representations of $N = 1$ SUSY in $d = 2 + 1$ dimensions, that consist of one real boson field and one real two-component spinor.

Since SUSY is not broken completely we can try to construct the Lagrangian, where the remnant supersymmetry is manifest not only for quadratic fluctuations. We want to take into account interaction between different modes on the wall, and make sure that it is indeed supersymmetric. To this end we introduce three-dimensional superspace, which consists of two real Grassmann coordinates and usual $(2 + 1)$ Minkowski space. The superfield is constructed from one real boson and two component real Majorana spinor as follows,

$$\Phi = \phi + \sqrt{2}\theta\alpha + \theta\theta f \tag{130}$$

where $f$ is an auxiliary field. Covariant derivative is given by

$$D_\alpha = \frac{1}{2}\frac{\partial}{\partial \theta^\alpha} - i(\gamma_\mu \theta)_\alpha \partial^\mu \tag{131}$$

where $\gamma_\mu$ are two-by-two real Majorana matrices,

$$\gamma_0 = i\tau_2\,, \quad \gamma_1 = \tau_3\,, \quad \gamma_2 = \tau_1\,, \tag{132}$$

and $\tau$ are the Pauli matrices. The most general renormalizable Lagrangian that describes the interaction of two sets of superfields is

$$\mathcal{L} = \int d^3x d^2\theta \left\{ \sum_i \frac{1}{2}(D\Phi_i^A)^2 + \frac{1}{2}(D\Phi_i^B)^2 - m_i\,(\Phi_i^A\Phi_i^A - \Phi_i^B\Phi_i^B) \right.$$
$$\left. + \sum_{ijk} \frac{1}{3}B_{ijk}(\Phi_i^A\Phi_j^A\Phi_k^A + \Phi_i^B\Phi_j^B\Phi_k^B) + C_{ijk}(\Phi_i^A\Phi_j^B\Phi_k^B + \Phi_i^B\Phi_j^A\Phi_k^A) \right\}, \tag{133}$$

where

$$\Phi_i^A = \tilde{a}_i + \sqrt{2}\theta\alpha_i + \theta^2 f_i^A;$$
$$\Phi_i^B = \tilde{b}_i + \sqrt{2}\theta\beta_i + \theta^2 f_i^B. \tag{134}$$

Mass terms for fields $\Psi^A$ and $\Psi^B$ have a different sign in accord with Eq.(128). In $(2+1)$ dimensions the chiral transformations are absent and the sign of the fermion mass can not be made positive. With the aid of formulae given in Appendix B, we can cast the Lagrangian (6), after integration over $z$, in the form (133). This explicitly demonstrates, that the dynamics of the modes on the domain wall is supersymmetric. The absence of the corrections to the vacuum energy, inherent to the supersymmetric theories, ensures that the domain wall energy density is not renormalized, implying the domain wall to be the BPS-saturated state.

### C. Correction to the domain wall shape

While the domain wall energy density is fixed by its classical value, the shape of the solution is subject to corrections, regardless of the presence of SUSY in Lagrangian (133).

Indeed, the "tadpole" diagram, that arises from the cubic terms of Lagrangian (133), gives rise to the linear term in the effective Lagrangian. It shifts the equilibrium position of the fields and each mode thus acquires a vacuum expectation value, specific for every given mode. This leads to a change in the soliton shape,

$$\delta \phi_w(z) = \sum a_i(z) \frac{\phi_i^A + \phi_i^B}{\sqrt{2}} + \sum b_i(z) \frac{\phi_i^A - \phi_i^B}{\sqrt{2}} \tag{135}$$

where $a_i$ ($b_i$) is an eigenfunction of Eq.(124) and $\phi_i^A$ ($\phi_i^B$) is the vacuum expectation of the lowest component of the superfield $\Phi_i^A$ ($\Phi_i^B$). Once we established the existence of the "tadpole" graph, it is easier to compute it directly in four dimensions. The part of Lagrangian (6) with the cubic coupling has the form

$$\mathcal{L} = -\int d^4x \left\{ \mathcal{W}'''(\phi_w) \mathcal{W}''(\phi_w) a \left(a^2 + b^2\right) + \frac{1}{2} \mathcal{W}'''(\phi_w) \left[ a(\Psi^+\Psi^- + \Psi^-\Psi^+) + b(\Psi^-\Psi^- + \Psi^+\Psi^+) \right] \right\}. \tag{136}$$

To compute the tadpoles we need to know the propagators of the fields entering the Lagrangian (120), (136) in the domain wall background. At this point we may notice, that there is a partial cancelation between the boson and fermion loops. Indeed, the following relation holds between the traces of the bosonic and fermionic Green's functions,

$$\begin{aligned}
\text{Tr } G_{\Psi^+\Psi^-}[x,x] &= \frac{2\mathcal{W}''}{\partial'^2 + \mathcal{O}^2\mathcal{O}^1} = 4\,\mathcal{W}''(\phi_w)\, G_a[x,x], \\
\text{Tr } G_{\Psi^-\Psi^+}[x,x] &= \frac{2\mathcal{W}''}{\partial'^2 + \mathcal{O}^1\mathcal{O}^2} = 4\,\mathcal{W}''(\phi_w)\, G_b[x,x], \\
\text{Tr } G_{\Psi^+\Psi^+}[x,x] &= \text{Tr } G_{\Psi^-\Psi^-}[x,x] = 0,
\end{aligned} \tag{137}$$

where trace is taken over the Lorentz indices. This relation makes the cancelation transparent. Taking into account combinatorial factor 3 in the $a^3$ vertex, the total result is the difference between the contribution of loops with field $a$ and $b$ inside.

The corresponding Green's functions are easily found [12] and we end up with the following term in the effective Lagrangian,

$$\Delta \mathcal{L}^{eff} = -\int d^4x \left\{ 4\lambda^2 \phi_w(z)(G^a[x,x] - G^b[x,x]) \right\} a(x), \tag{138}$$

where $G^{a,b}[x,y]$ is the propagator of the field $a$ ($b$) in the background of the domain wall. The presence of the linear term in the Lagrangian shifts the equilibrium positions of the fields away from zero. To find a new vacuum configuration we must solve the equation of motion for the boson field $a$, and, thus, obtain the corrected shape $\tilde{\phi}_w$ of the domain wall,

$$\tilde{\phi}_w = \phi_w(\bar{\lambda},\bar{m}) + \left\{\frac{\lambda^2}{16\pi\sqrt{3}}\right\}\phi_w(1-[\frac{\lambda}{m}\phi_w]^2), \tag{139}$$

where

$$\bar{\lambda} = \lambda\left(1-\lambda^2\left[\frac{3}{8\pi^2}\ln(M_0) - \frac{1}{8\pi\sqrt{3}}\right]\right), \quad \bar{m} = m\left(1-\lambda^2\left[\frac{1}{4\pi^2}\ln(M_0) - \frac{1}{8\pi\sqrt{3}}\right]\right)$$

are the renormalized coupling constant and mass, and $M_0$ is the ultraviolet cutoff in units of $m$. The classical solution falls off at infinity as $e^{-2\bar{m}\,z}$, implying that a one-particle state has a mass $\bar{m}$. Thus, $\bar{m}$ is the mass renormalized on the mass-shell.

Note that the second term in Eq.(139) can be represented as

$$-\frac{1}{32\pi\sqrt{3}}\frac{\lambda^2}{m^2}\frac{\partial\phi_w}{\partial z^2}. \tag{140}$$

Thus, the change in the wall shape corresponds to a second order effect in the shift of the wall center.

The presence of the corrections to the shape of the domain wall has an impact on the low energy dynamics of the fields on the wall. The operator $\mathcal{O}^2\mathcal{O}^1(z)$ is known to have a zero mode, corresponding to the breaking of the translational symmetry by the domain wall [30]. Exactly one bosonic and one (two component) fermionic zero modes exist, which reflects the balance of degrees of freedom in SUSY theories (the operator $\mathcal{O}^1\mathcal{O}^2(z)$ does not have zero modes). This provides a zero mode superfield in the Lagrangian (133) denoted by $\Phi_0$. This superfield (see Eqs. (B4), (B5)) does not have self-interaction, if all non-zero modes are omitted the Lagrangian for the zero mode superfield has the form

$$\mathcal{L}_{ZM} = \int d^3x d^2\theta \frac{1}{2}(D\Phi_0)^2. \tag{141}$$

The zero mode, however, can mix with the non-zero modes through trilinear coupling and, thus, acquire a mass term induced by loop corrections with the massive particles inside the loop. This is what is to be expected since we established that the shape of wall does experience corrections. The zero mode, being obtained by applying the broken symmetry generator ($\partial_z$ in our case) to the domain wall solution, also receives corrections. After we incorporate the loop effects into the effective Lagrangian, it is a certain linear combination of all modes that remains massless and represents the Goldstone mode of the broken translational invariance.

## V. APPLICATIONS

In this section we will consider some obvious applications of the ideas that were discussed above.

### A. Tunneling in non-supersymmetric non-relativistic quantum mechanics

One of the most interesting results is the exact expression for $\varepsilon$ which can be obtained even in the absence of the explicit solution for the domain wall. This technique can be used in complicated multidimensional quantal problems with tunneling, for evaluating the tunneling exponents, in the situations where the WKB methods are inefficient, and the least action trajectories are not obvious. It is well-known, that in these cases one has usually resort to numerical solutions, which are quite complicated – since the effect sought for is exponentially small, achieving high accuracy is a non-trivial task.

Exploiting the methods developed here the tunneling exponential in conventional (non-supersymmetric) quantum mechanics can be analytically obtained even without exact knowledge of the tunneling trajectory.

Below we will consider a specific example. It is quite clear however, that the situation is general. Consider four-dimensional (non-supersymmetric) quantum mechanics with the potential

$$V = |-\alpha + \gamma(X - \Phi)^2|^2 + |\beta\Phi - \gamma(X - \Phi)^2|^2 \qquad (142)$$

where the four degrees of freedom in question are the real and imaginary parts of $X$ and $\Phi$. All particle masses are put to 2. This is a reduction of the model discussed in Sect. III.B, see Eq. (95). The potential is horrible, it has two degenerate minima where the classical potential energy vanishes. We are interested in the probability of the system to tunnel from one minimum to another. As well-known, this probability is equal to the action of the classical trajectory connecting two minima in the Euclidean time. After the Euclidean rotation the equation of motion exactly coincides with that for the wall, what was $z$ in the case of the wall becomes the Euclidean time, and what was $\varepsilon$ becomes the classical action. Since the solution is BPS-saturated the value of the action can be immediately read off from Eq. (96), without any calculations. The tunneling probability is

$$\propto \exp(-|8\alpha^{3/2}3^{-1}\gamma^{-1/2}|)\,.$$

### B. Multiparticle production amplitudes at the kinematical threshold

Non-vanishing corrections to the shape of the domain wall (Eq.(139)) have fenomenological consequences. As was shown in Ref. [10], the multiparticle production amplitudes at the kinematical threshold are related to the shape of the classical solution existing in the theory. We briefly review the main arguments leaving all the details aside.

In the presence of the external source $\rho(x)$ the multiparticle threshold amplitude at the tree-level can be written as

$$\mathcal{A}_n \equiv \langle n|\phi(x)|0\rangle = \prod_{a=1}^{n} \int (d^4 x_a) e^{-i\omega t_a}(\omega^2 - m^2)\frac{\partial}{\partial \rho(t_a)}\phi_{cl}(t)\,|_{\rho=0} \qquad (143)$$

where $\phi_{cl}$ is the solution of the classical equations of motion in the presence of the *external source* $\rho(t)$ and we made use of the reduction formula for the $n$-particle state with zero spatial momenta $p_a = (\omega, 0, 0, 0,)$. Since we are dealing with the threshold amplitudes, both $\rho$ and $\phi$ can be chosen to depend upon time $t$ only, i.e. $\rho = \rho(t)$ and $\phi = \phi(t)$. The key point is that $\phi_{cl}$ can be shown to have the following expansion in powers of coupling constant $\lambda$ (cf. Eq.(104))

$$\phi_{cl} = z(t) \left(1 + \sum_n c_n (\lambda\, z(t))^n\right) \tag{144}$$

where

$$z(t) \equiv \frac{\rho}{\omega^2 - 4m^2} e^{i\omega t}. \tag{145}$$

Factor $4m^2$ appears, since with our choice of the superpotential (104), fluctuations over the vacuum have the mass $2m$. Taking the limit as $\rho \to 0$, $\omega \to 2m$ and keeping $z(t)$ finite we obtain the threshold amplitude in the form

$$\langle n|\phi(x)|0\rangle = \left(\frac{\partial}{\partial z(t)}\right)^n \phi_{cl}(z(t))\,|_{z=0}, \tag{146}$$

and now $\phi_{cl}$ is the solution of the (static) classical equations of motion *without the source*

$$\frac{\partial^2}{\partial t^2}\phi_{cl}(t) = -(\mathcal{W}')^2(\phi_{cl}(t)), \tag{147}$$

with the boundary condition

$$\phi_{cl}(t) \to z(t) \quad \text{as} \quad \lambda \to 0. \tag{148}$$

The solution of Eq.(147) can be obtained from the domain wall solution by substituting $z \to it$. Rewriting it in terms of $z(t)$ with the proper boundary conditions we obtain [10]

$$\phi_{cl} = \frac{m}{\lambda} \frac{1 + \lambda z(t)/(2m)}{1 - \lambda z(t)/(2m)} \tag{149}$$

The loop corrections to the amplitudes $\mathcal{A}_n$ are, in turn, given by the quantum corrections to the shape of the classical solution [12,13], which are given in Eq.(139). All the details of the calculation can be found in Refs. [12,13], where non-supersymmetric theories were considered. Here we merely present the final result

$$\mathcal{A}_n \equiv \langle n|\phi(0)|0\rangle = n! \left(\frac{\bar{\lambda}}{2\bar{m}}\right)^{n-1} \left(1 + n(n-1)\frac{\lambda^2}{8\pi\sqrt{3}}\right). \tag{150}$$

Note, that the quantum corrections to the amplitudes $\mathcal{A}_n$, are by factor $1/3$ smaller than ones in non-supersymmetric theory [13]. The result was known to M. Voloshin (private communication).

## VI. "KINK" IN 1 PLUS 1 DIMENSIONS

Now we turn to another example of supersymmetric theory, with SUSY being partially broken in the soliton sector. The two-dimensional analog of the Wess-Zumino model with the minimal ($N = 1$) supersymmetry was analyzed long ago [5]. The model is similar to the four-dimensional example. The theory possesses a classical solution – the "kink", that breaks 1/2 of supersymmetry. However, contrary to the domain walls in four dimensions, the mass of the "kink" does receive a perturbative correction.

The Lagrangian of the two-dimensional analog of the Wess-Zumino model describes the interaction of a real scalar field $\phi$ and a real two component spinor $\Psi$. It reads, in components,

$$\mathcal{L} = \int d^2x \left\{ \frac{1}{2}(\partial_\mu \phi)^2 - \mathcal{W}'(\phi)^2 + \frac{1}{2}\bar{\Psi}i\hat{\partial}\Psi - \frac{1}{2}\mathcal{W}''(\phi)\bar{\Psi}\Psi \right\} \tag{151}$$

where $\mathcal{W}(\phi)$ is the same as in Eq.(104) and

$$\gamma^0 = \sigma^2, \quad \gamma^1 = i\sigma^3.$$

The consideration goes essentially along the same lines as in the previous section: we expand the Lagrangian around the classical background and investigate the dynamics of the quantum fluctuations. There are few changes, however. Now, there are only two supersymmetry generators $Q_1$ and $Q_2$. The "kink" solution (109) breaks 1/2 of SUSY [31], and only the generator $Q_2$ survives. The surviving generator annihilates the "kink" solution, while applying the broken generator $Q_1$ to the "kink" will produce a fermionic zero mode, thus, transforming the "kink" into the left-moving fermion, and the "antikink" into the right-moving one.

Let us note, that in contrast to the four-dimensional case, supersymmetry of the theory is not spontaneously broken. The "kink" solution has finite energy and can be created from the vacuum and destroyed as a result of a field fluctuation. In this sense the "kink" solution is a one-particle state, rather than the vacuum state of the theory, as it is the case with the domain wall solution of the previous section. The "kink" can be pair-produced. Supersymmetry algebra will be realized linearly on the states with different number of "kinks" and "antikinks". We are interested, however, not in the general supersymmetry of the theory, but in the question how it is realized in the *given* soliton background. In the given background $Q_1$ must be considered as spontaneously broken.

The question of the corrections to the mass of the "kink" was addresses more than once in the literature (see [31], [33], and references therein). There are two basic approaches that differ in the way the boundary conditions on the quantum fluctuations around the "kink" solution are imposed.

In the first approach, that was taken in [31], non-supersymmetric boundary conditions were imposed at spatial infinities. Since supersymmetry is explicitly broken by the choice of the boundary condition, the volume effects associated with the production of virtual fermions and bosons will introduce a non-vanishing mass correction.

To see this let us look at the following eigenvalue equations (cf. Eq.(121), (122)) that can be easily obtained by expanding (151) around the "kink" up to the quadratic terms :

$$\mathcal{O}^2\mathcal{O}^1(z)\phi_n = \lambda_n^{+2}\phi_n ,$$
$$\mathcal{O}^2\mathcal{O}^1(z)\Psi_n^+ = \lambda_n^{+2}\Psi_n^+ , \qquad \mathcal{O}^1\mathcal{O}^2(z)\Psi^- = \lambda_n^{-2}\Psi_n^- \qquad (152)$$

where

$$i\gamma^1\Psi^\pm = \pm\Psi^\pm.$$

(Note that $\Psi^\pm$ are real one-component fields in contrast to the fields $\Psi_\alpha^\pm$ defined in the previous section that had one independent complex component. With our choice of $\gamma$ matrices $\Psi^- = \Psi_1$ and $\Psi^+ = \Psi_2$ .)

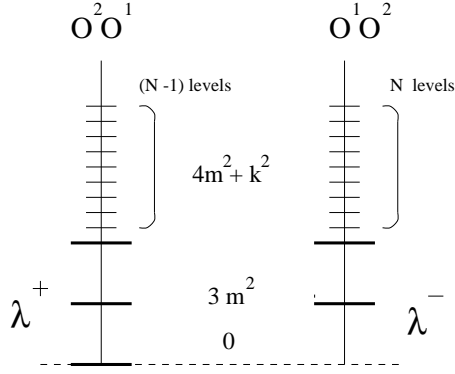

FIG. 2. Eigenvalues of two operators.

If non-supersymmetric boundary conditions

$$\begin{cases} \Psi^{\pm}(-L) = \Psi^{\pm}(L) \\ \phi(-L) = \phi(L) \\ \phi'(-L) = \phi'(L) \end{cases} \quad . \tag{153}$$

are imposed ($L \to \infty$), then in the limit of infinite volume two operators $\mathcal{O}^1\mathcal{O}^2$ and $\mathcal{O}^2\mathcal{O}^1$ will have different density of the eigenstates. This difference can be readily deduced from the asymptotic form of the eigenfunctions of two operators and is equal to [30,31]

$$\frac{d\,n^+}{d\,k} - \frac{d\,n^-}{d\,k} = \frac{2}{\pi}\frac{1}{4+k^2} \tag{154}$$

where $k$ is the spatial momentum in units of $m$. Integrating over the spatial momentum we obtain,

$$N^+ - N^- = 1 \tag{155}$$

where $N^+$ ($N_-$) is the number of the continuous eigenstates of the operator $\mathcal{O}^2\mathcal{O}^1$ (or $\mathcal{O}^1\mathcal{O}^2$). This difference in the density of the fermionic and bosonic eigenstates will produce a non-zero mass correction that was derived in [31].

An alternative logical step would be to impose a supersymmetric boundary conditions. Then, the consideration will run in parallel to what was done with the domain walls in the previous sections. This approach was undertaken in [33] where the supersymmetric boundary conditions suggested in [32] were modified to be

$$\begin{cases} \phi(-L) = \phi(L) \\ \mathcal{O}^1\phi(-L) = \mathcal{O}^1\phi(L) \\ same \ \ as \ \ before \ \ for \ \ fermions \end{cases} . \qquad (156)$$

Then, all volume contributions in the one-loop correction cancel out, the boson correction is canceled by the fermion one. Mass renormalization, nevertheless, still occurs in [33] due to a surface term that appears after integrating the term $(\partial_\mu \phi)^2$ in Lagrangian (151) by parts and using (156).

Let us note, that the cancelation of the bulk effects is usual for the theories with unbroken supersymmetry, the same situation occurred when domain walls were considered. However, the appearance of the boundary correction is specific for supersymmetric theories in two dimensions, where superspace can not be divided into chiral subspaces. In four dimensions, quantum corrections can not induce non-vanishing $\langle \Phi^n \rangle$ for the chiral field $\Phi$ at the boundary, hence no boundary correction can appear (in other words the superpotential is not renormalized in four dimensions). In contrary to four-dimensional case, in two dimensional models with the minimal supersymmetry no chiral field exists, and, thus, $\langle \Phi^n \rangle$ can have non-zero value, inducing non-vanishing boundary terms. In two-dimensional models with the extended supersymmetry ($N = 2$) the quantum corrections from the boundaries are absent much in the same way as in the domain wall problem, and the soliton mass remains unrenormalized.

There is a subtlety in imposing the supersymmetric boundary conditions. The point is that the soliton mass correction is given by the surface term, i.e. it is due to the region most affected by the choice of the boundary condition. This is true for any supersymmetric choice of the boundary conditions. The answer is saturated on the boundaries and, thus, largely affected by particular choices of the boundary conditions. It is necessary that the given set of boundary conditions corresponds to a pure soliton state rather than some excited soliton state. The following simple argument illustrates that this is not the case with the boundary conditions (156). Indeed, Eq.(156) implies that the expectation value of the operator $\langle \bar{\Psi}\Psi \rangle$ is the same at $\pm\infty$. On the other hand, since the fermion mass changes its sign as one goes from plus to minus infinity, the expectation value of $\bar{\Psi}\Psi$ in the pure soliton sector should also change sign. This inconsistency implies, that the boundary conditions (156) project out an excited soliton state containing a fermion pair. Physically, it is clear that the problem of calculation of the soliton mass must be formulated in a way independent of the specific choice of the boundary conditions. This can be done by considering a local form of the BPS saturation condition. For more details see [39].

Let us note, that in line with the arguments presented above, the subtlety with the boundary conditions is immaterial in the four-dimensional theories, and, thus, need not be considered when dealing with the domain walls. In four dimensions the expectation value of the operators $\langle\Phi^n\rangle$ is zero due to the same argument that leads to the non-renormalization of the superpotential in four dimensions. Thus, the contamination of the soliton does not occur, and the particular choice of the boundary conditions does not affect the solution.

## VII. CONCLUSIONS

The existence of the central extension in the $N=1$ superalgebra in four-dimensional gauge theories was first observed in Ref. [1,2]. We present a detailed derivation of this extension in the general case: non-Abelian gauge theories with a superpotential. Our derivation takes advantage of the well-established "geometric" anomalies, such as the anomaly in the trace of the energy-momentum tensor. In fact, we show that the latter is in one-to-one correspondence with the central extension in the $N=1$ superalgebra which we thoroughly discuss here.

In few cases when the diagrammatic computation is simple, the result agrees with the general answer obtained indirectly. The direct diagrammatic derivation for the general case is the subject of further work.

The fact that the superalgebra is extended and the central charge is non-zero leads to far-reaching consequences: the existence of the BPS-saturated topological defects. In-depth studies of the BPS-saturated domain walls is carried out. We derive the creek equations and find necessary conditions for the solutions to exist, and other useful criteria. BPS saturation means that solutions partially preserve supersymmetry, 1/2 of SUSY remains unbroken in the given background. This, in turn, entails *exact calculability* of the domain wall energy density. It is related to the central charge for the corresponding topological defect. The domain wall energy density receives no quantum corrections.

The shape of the domain walls can not be exactly deduced from the central extension and is subject to perturbative loop corrections even in the presence of residual supersymmetry.

We discuss in great detail the generalized Wess-Zumino models. Apart from the domain walls, string-like topological defects, that are BPS-saturated, are shown to exist. This is a novel step, since the BPS-saturated strings were not discussed in Ref. [1]. It is clear that consideration of the BPS-saturated topological solitons in the supersymmetric ($N = 1$) Yang-Mills theories only begins, and will be continued further, see also Ref. [22].

In the minimal Wess-Zumino model, where the solutions of the classical equations of motion are known analytically, all conclusions based on the analysis of the central extension can be explicitly checked. In the strongly interacting gauge theories, to explicitly reveal the way the central extension works, one has to resort to effective Lagrangians of the type suggested by Veneziano and Yankielowicz, as was done in [22]. We do not rule out that other, more elegant methods will eventually emerge.

*Note Added.* A few days ago, when the write up of this paper was being completed, Witten's work [37] was submitted to hep-th, where constructions similar to those discussed above were discussed in the context of the $D$-brane-based approach to QCD.

*Acknowledgments.* The authors are grateful to G. Dvali, I. Kogan, A. Kovner, A. Smilga and M. Voloshin for numerous useful discussions at various stages of this work. A few final touches were added to the paper after the work had been reported at the Workshop on *Non-Perturbative Aspects of Quantum Field Theory*, Isaac Newton Institute for Mathematical Science, University of Cambridge. M.S. is grateful to the participants of the Workshop, and especially to M. Green and P. Van Baal, for stimulating questions.

We would like to thank S.-J. Rey and P. Townsend for bringing Refs. [15,16] to our attention.

This work was supported in part by DOE under the Grant number DE-FG02-94ER40823.

# APPENDIX A: WESS-ZUMINO MODEL AND SUPERSYMMETRIC GLUODYNAMICS IN SPINORIAL FORMALISM

## 1. Notation

We first specify our notations and conventions. They are close but not identical to these of Bagger and Wess [34]. The main distinction is that we choose the metric to be $(+---)$. There are also distinctions in normalization, see Eq.(A19). The left-handed spinor is denoted by undotted indices, e.g. $\eta_\beta$. The right-handed spinor is denoted by dotted indices, e.g. $\bar{\xi}^{\dot\beta}$. (This convention is standard in supersymmetry but is opposite to one accepted in the textbook [35]). The Dirac spinor $\Psi$ then takes the form

$$\Psi = \begin{pmatrix} \bar{\xi}^{\dot\beta} \\ \eta_\beta \end{pmatrix}. \tag{A1}$$

Lowering and raising of the spinorial indices is done by multiplying by the Levi-Civita tensor from the left,

$$\chi^\alpha = \epsilon^{\alpha\beta}\chi_\beta \,, \quad \chi_\alpha = \epsilon_{\alpha\beta}\chi^\beta, \tag{A2}$$

and the same for the dotted indices, where

$$\epsilon^{\alpha\beta} = -\epsilon^{\beta\alpha} \,, \quad \epsilon^{12} = -\epsilon_{12} = 1. \tag{A3}$$

The products of the undotted and dotted spinors are defined as follows:

$$\eta\chi = \eta^\alpha\chi_\alpha = -\eta_\alpha\chi^\alpha \,, \quad \bar\eta\bar\chi = \bar\eta_{\dot\alpha}\bar\chi^{\dot\alpha}. \tag{A4}$$

Under this convention $(\eta\chi)^+ = \bar{\chi}\bar{\eta}$. Moreover

$$\theta^\alpha\theta^\beta = -\frac{1}{2}\epsilon^{\alpha\beta}\theta^2, \quad \theta_\alpha\theta_\beta = \frac{1}{2}\epsilon_{\alpha\beta}\theta^2,$$
$$\bar{\theta}^{\dot\alpha}\bar{\theta}^{\dot\beta} = \frac{1}{2}\epsilon^{\dot\alpha\dot\beta}\bar{\theta}^2, \quad \bar{\theta}_{\dot\alpha}\bar{\theta}_{\dot\beta} = -\frac{1}{2}\epsilon_{\dot\alpha\dot\beta}\bar{\theta}^2. \tag{A5}$$

The vector quantities (representation $(\frac{1}{2}, \frac{1}{2})$) are obtained in the spinorial formulation by multiplication by

$$(\sigma^\mu)_{\alpha\dot\beta} = \{1, \vec{\tau}\}_{\alpha\dot\beta} \tag{A6}$$

where $\vec{\tau}$ stands for the Pauli matrices, for instance,

$$A_{\alpha\dot\beta} = A_\mu(\sigma^\mu)_{\alpha\dot\beta}. \tag{A7}$$

Note that

$$A_\mu B^\mu = \frac{1}{2}A_{\alpha\dot\beta}B^{\alpha\dot\beta}, \quad A_{\alpha\dot\beta}A^{\gamma\dot\beta} = \delta_\alpha^\gamma\, A_\mu A^\mu. \tag{A8}$$

The square of the four-vector is understood as

$$A^2 = A_\mu A^\mu = \frac{1}{2}A_{\alpha\dot\beta}A^{\alpha\dot\beta}. \tag{A9}$$

If the matrix $(\sigma^\mu)_{\alpha\dot\beta}$ is "right-handed" it is convenient to introduce its "left-handed" counterpart,

$$(\bar{\sigma}^\mu)^{\dot\beta\alpha} = \{1, -\vec{\tau}\}_{\dot\beta\alpha}. \tag{A10}$$

The matrices that appear in dealing with representations $(1, 0)$ and $(0, 1)$ are

$$(\vec{\sigma})^\alpha_\beta = \vec{\tau}_{\alpha\beta}, \quad (\vec{\sigma})^{\alpha\beta} = \epsilon^{\beta\delta}\vec{\sigma}^\alpha_\delta \tag{A11}$$

and the same for the dotted indices. The matrices $(\vec{\sigma})^{\alpha\beta}$ are symmetric, $(\vec{\sigma})^{\alpha\beta} = (\vec{\sigma})^{\beta\alpha}$. In explicit form

$$(\vec{\sigma})^{\dot\alpha\dot\beta} = \{\tau^3, -i\mathbf{1}, -\tau^1\}_{\alpha\beta}; \quad (\vec{\sigma})^{\alpha\beta} = \{\tau^3, i\mathbf{1}, -\tau^1\}_{\alpha\beta}. \tag{A12}$$

Note that with our definition

$$(\vec{\sigma})_{\dot{\alpha}\dot{\beta}} = \{\,-\tau^3\,,\,-i\mathbf{1}\,,\,\tau^1\,\}_{\alpha\beta};\quad (\vec{\sigma})_{\alpha\beta} = \{\,-\tau^3\,,\,i\mathbf{1}\,,\,\tau^1\,\}_{\alpha\beta}\,.$$

The left (right) coordinates $x_{L,R}$ and covariant derivatives are

$$(x_L)_{\alpha\dot{\alpha}} = x_{\alpha\dot{\alpha}} - 2\,i\,\theta_\alpha\bar{\theta}_{\dot{\alpha}}\,,\quad (x_R)_{\alpha\dot{\alpha}} = x_{\alpha\dot{\alpha}} + 2\,i\,\theta_\alpha\bar{\theta}_{\dot{\alpha}}\,,$$
$$D_\alpha = \frac{\partial}{\partial\theta^\alpha} - i\,\partial_{\alpha\dot{\alpha}}\bar{\theta}^{\dot{\alpha}}\,,\quad \bar{D}_{\dot{\alpha}} = -\frac{\partial}{\partial\bar{\theta}^{\dot{\alpha}}} + i\,\theta^\alpha\partial_{\alpha\dot{\alpha}}, \tag{A13}$$

so that

$$\{D_\alpha\,\bar{D}_{\dot{\alpha}}\} = 2i\partial_{\alpha\dot{\alpha}}\,,$$
$$\bar{D}_{\dot{\beta}}(x_L)_{\alpha\dot{\alpha}} = 0\,,\quad D_\beta(x_R)_{\alpha\dot{\alpha}} = 0\,,$$
$$\bar{D}_{\dot{\beta}}(x_R)_{\alpha\dot{\alpha}} = -4\,i\,\theta_\alpha\epsilon_{\dot{\beta}\dot{\alpha}}\,,\quad D_\beta(x_L)_{\alpha\dot{\alpha}} = 4\,i\,\bar{\theta}_{\dot{\alpha}}\epsilon_{\beta\alpha}. \tag{A14}$$

The law of the supertranslation is

$$\theta \to \theta + \epsilon\,,\quad \bar{\theta} \to \bar{\theta} + \bar{\epsilon}\,,$$
$$x_{\alpha\dot{\beta}} \to x_{\alpha\dot{\beta}} - 2\,i\,\theta_\alpha\bar{\epsilon}_{\dot{\beta}} + 2\,i\,\epsilon_\alpha\bar{\theta}_{\dot{\beta}}\,,$$
$$(x_L)_{\alpha\dot{\beta}} \to (x_L)_{\alpha\dot{\beta}} - 4\,i\,\theta_\alpha\bar{\epsilon}_{\dot{\beta}}\,,$$
$$(x_R)_{\alpha\dot{\beta}} \to (x_R)_{\alpha\dot{\beta}} + 4\,i\,\epsilon_\alpha\bar{\theta}_{\dot{\beta}}. \tag{A15}$$

It corresponds to the infinitesimal transformation of the superfield in the form

$$\delta\Phi = i\left(\epsilon Q + \bar{\epsilon}\bar{Q}\right)\Phi \tag{A16}$$

where

$$Q_\alpha = -i\frac{\partial}{\partial\theta^\alpha} + \partial_{\alpha\dot{\alpha}}\bar{\theta}^{\dot{\alpha}},$$
$$\bar{Q}_{\dot{\alpha}} = i\frac{\partial}{\partial\bar{\theta}^{\dot{\alpha}}} - \theta^\alpha\partial_{\alpha\dot{\alpha}}. \tag{A17}$$

The integrals over the Grassmann variable are normalized as follows

$$\int d^2\theta\,\theta^2 = 2\,,\quad \int d^4\theta\,\theta^2\,\bar{\theta}^2 = 4\,, \tag{A18}$$

and we define

$$\{\ldots\}_F = \frac{1}{2}\int d^2\theta\{\ldots\}\,,\quad \{\ldots\}_D = \frac{1}{4}\int d^4\theta\{\ldots\}. \tag{A19}$$

## 2. Wess-Zumino model

The minimal Wess-Zumino model describes (self) interaction of the chiral superfield

$$\Phi(x_L, \theta) = \phi(x_L) + \sqrt{2}\theta\psi(x_L) + \theta^2 F. \tag{A20}$$

The action is given in Eq. (6). In components it has the form,

$$\mathcal{L} = (\partial^\mu \phi^+)(\partial^\mu \phi) + \psi^\alpha i \partial_{\alpha\dot\alpha} \bar\psi^{\dot\alpha} + F^+ F$$
$$+ \left\{ F\, \mathcal{W}'(\phi) - \frac{1}{2} \mathcal{W}''(\phi)\, \psi\psi + H.C. \right\}. \tag{A21}$$

The model possesses a supermultiplet of currents,

$$J_{\alpha\dot\alpha} = \frac{1}{6}\left\{ D_\alpha \Phi \bar D_{\dot\alpha} \bar\Phi - 2\Phi \partial_{\alpha\dot\alpha} \bar\Phi + 2\bar\Phi i \partial_{\alpha\dot\alpha} \Phi \right\}. \tag{A22}$$

The $\theta$ component of the supermultiplet is related (cf. Eq. (A38)) to the supercurrent, that has the form

$$J_{\alpha\beta\dot\beta} = 2\sqrt{2}\left\{ \left[(\partial_{\alpha\dot\beta}\phi^+)\psi_\beta - i\,\epsilon_{\beta\alpha} F \bar\psi_{\dot\beta}\right] - \right.$$
$$\left. \frac{1}{6}\left[\partial_{\alpha\dot\beta}(\psi_\beta\phi^+) + \partial_{\beta\dot\beta}(\psi_\alpha\phi^+) - 3\epsilon_{\beta\alpha}\partial_{\dot\beta}^\gamma(\psi_\gamma\phi^+)\right]\right\} \tag{A23}$$

where

$$F = -\frac{\partial \bar{\mathcal{W}}}{\partial \phi^+}, \tag{A24}$$

and $\mathcal{W}$ is the superpotential. In the spin-vector form the supercurrent is

$$J_\alpha^\mu = \frac{1}{2}(\bar\sigma^\mu)^{\dot\beta\beta} J_{\alpha\beta\dot\beta}. \tag{A25}$$

This current is conserved,

$$\partial_\mu J_\alpha^\mu \equiv \frac{1}{2}\partial^{\dot\beta\beta} J_{\alpha\beta\dot\beta} = 0. \tag{A26}$$

The supercharge is defined

$$Q_\alpha = \int d^3x\, J_\alpha^0. \tag{A27}$$

Terms in the second square brackets are the total spatial derivatives and, thus, do not contribute to the supercharge. These terms have to be added if we require the supercurrent to enter the same supermultiplet as the energy-momentum tensor. Then at the tree level, the current has vanishing trace $J^\alpha_{\alpha\dot\alpha}$ in the the theory, that is classically conformally invariant. The relative coefficient in front of the second bracket is most readily fixed by the requirement that $\epsilon^{\alpha\beta} J_{\alpha\beta\dot\beta}$ reduces to the equation of motion.

The anticommutator of $Q_\alpha$ and $\bar{Q}_{\dot\beta}$ is standard,

$$\{Q_\alpha \, \bar{Q}_{\dot\beta}\} = 2\, P_{\alpha\dot\beta}\,, \tag{A28}$$

while that of $Q_\alpha$ and $Q_\beta$ contains a central extension

$$\{Q_\alpha \, Q_\beta\} = (-4i)(\vec\sigma)_{\alpha\beta} \int d^3x \vec\nabla \left\{ \bar{\mathcal{W}} - \frac{1}{3}\bar\Phi \frac{\partial\bar{\mathcal{W}}}{\partial\bar\Phi} \right\}. \tag{A29}$$

At the loop level, due to the quantum anomaly, an additional term on the right-hand side appears; being combined with $(1/3)\bar\Phi\partial\bar{\mathcal{W}}/\partial\bar\Phi$, this additional term is a total superderivative, that gives no contribution to the central charge. Hence, we arrive at the following *exact* relation

$$\{Q_\alpha \, Q_\beta\} = (-4i)(\vec\sigma)_{\alpha\beta} \int d^3x \, \vec\nabla \bar{\mathcal{W}}\,. \tag{A30}$$

### 3. Supersymmetric gluodynamics

This model (as well as its $N = 2$ super generalization) was first introduced by Ferrara and Zumino [36]. The superfield $W_\alpha$ which includes the gluon strength tensor is

$$W_\alpha = \frac{1}{8}\, \bar{D}^2 \left( \mathrm{e}^{-V}\, D_\alpha\, \mathrm{e}^V \right) \tag{A31}$$

where $V$ is the vector superfield; in Wess-Zumino gauge

$$V = -2\theta^\alpha \bar\theta^{\dot\alpha} A_{\alpha\dot\alpha} - 2i\bar\theta^2(\theta\lambda) + 2i\theta^2(\bar\theta\bar\lambda) + \theta^2\bar\theta^2 D, \tag{A32}$$

$V = V^a T^a$ and $T^a$ stands for the generators of the gauge group $G$,

$$\mathrm{Tr}\left(T^a T^b\right) = \frac{1}{2}\delta^{ab}$$

for the fundamental representation. If the matter field is denoted by $S$, the supergauge transformation has the form

$$S \to e^{i\Lambda} S, \quad e^V \to e^{i\bar{\Lambda}} e^V e^{-i\Lambda}, \quad W_\alpha \to e^{i\Lambda} W_\alpha e^{-i\Lambda} \tag{A33}$$

where $\Lambda$ is an arbitrary chiral superfield ($\bar{\Lambda}$ is antichiral). In components

$$W_\alpha = i\left(\lambda_\alpha + i\theta_\alpha D - \theta^\beta G_{\alpha\beta} - i\theta^2 \mathcal{D}_{\alpha\dot\alpha}\bar\lambda^{\dot\alpha}\right) \tag{A34}$$

where $\lambda_\alpha$ is the gluino (Weyl) field, $\mathcal{D}_{\alpha\dot\alpha}$ is the covariant derivative, and $G_{\alpha\beta}$ is the gluon field strength tensor in the spinorial notation.

The standard gluon field strength tensor transforms as $(1,0) + (0,1)$ with respect to the Lorentz group. Projecting out pure $(1,0)$ is achieved by virtue of the $(\sigma)_{\alpha\dot\beta}$ matrices,

$$G_{\alpha\beta} = -\frac{1}{2} G_{\mu\nu}(\sigma^\mu)_{\alpha\dot\beta}(\sigma^\nu)_{\beta\dot\delta}\epsilon^{\dot\beta\dot\delta} = (\vec E - i\vec B)(\vec\sigma)_{\alpha\beta}. \tag{A35}$$

Then

$$G^{\alpha\beta} G_{\alpha\beta} = 2(\vec B^2 - \vec E^2 + 2i\vec E \vec B) = G_{\mu\nu} G^{\mu\nu} - i G_{\mu\nu}\tilde G^{\mu\nu}$$

where

$$\tilde G_{\mu\nu} = \frac{1}{2}\epsilon_{\mu\nu\alpha\beta} G^{\alpha\beta}, \quad (\epsilon_{0123} = -1). \tag{A36}$$

The Lagrangian of the supersymmetric gluodynamics is given by

$$\mathcal{L} = \left\{\frac{1}{4g_0^2}\mathrm{Tr}\int d^2\theta\, W^2 + H.C.\right\}, \quad \frac{1}{g^2} = \frac{1}{g^2} - \frac{i\vartheta}{8\pi^2}, \tag{A37}$$

where $\vartheta$ is the vacuum angle.

The supercurrent superfield takes the form (the trace is taken over the color indices only)

$$J_{\alpha\dot\alpha} = -\frac{2}{g^2}\mathrm{Tr}\left(W_\alpha e^{-V}\bar W_{\dot\alpha}\, e^V\right) =$$
$$-\frac{1}{g^2}\lambda_\alpha^a \bar\lambda_{\dot\alpha}^a - \frac{1}{2}\left\{i\theta^\beta\left(J_{\beta\alpha\dot\alpha} - \frac{2}{3}\epsilon_{\beta\alpha}\epsilon^{\gamma\delta} J_{\delta\gamma\dot\alpha}\right) + H.C.\right\} -$$
$$\theta^\beta\bar\theta^{\dot\beta}\left(J_{\alpha\dot\alpha\beta\dot\beta} - \frac{1}{3}\epsilon_{\alpha\beta}\epsilon_{\dot\alpha\dot\beta}\epsilon^{\gamma\delta}\epsilon^{\dot\gamma\dot\delta} J_{\gamma\dot\gamma\delta\dot\delta}\right) + \ldots \tag{A38}$$

where $J_{\beta\alpha\dot\alpha}$ is the supercurrent

$$J_{\beta\alpha\dot\alpha} = \frac{4i}{g^2}\mathrm{Tr}\left\{G_{\alpha\beta}\bar\lambda_{\dot\alpha}\right\} \tag{A39}$$

and $J_{\alpha\dot\alpha\beta\dot\beta}$ is the energy-momentum tensor,

$$J_{\alpha\dot\alpha\beta\dot\beta} = \frac{2}{g^2}\mathrm{Tr}\left\{G_{\alpha\beta}\bar G_{\dot\alpha\dot\beta} - i\,\lambda_\alpha \mathcal{D}_{\beta\dot\beta}\bar\lambda_{\dot\alpha} - i\,(\mathcal{D}_{\beta\dot\beta}\lambda_\alpha)\bar\lambda_{\dot\alpha}\right\} \tag{A40}$$

The energy-momentum tensor in the spinorial notation is related to $\theta_{\mu\nu}$ in the following way

$$J_{\alpha\dot\alpha\beta\dot\beta} = -(\sigma^i)_{\alpha\beta}(\sigma^j)_{\dot\alpha\dot\beta}\left\{\theta^{ij} + \theta^{00}g^{ij} - \epsilon^{ijk}\theta^{0k}\right\} + \frac{1}{2}\epsilon_{\alpha\beta}\epsilon_{\dot\alpha\dot\beta}\theta^\mu_\mu \tag{A41}$$

where $i,j,k = 1,2,3$, $g_{\mu\nu}$ is the metric tensor and matrices $(\sigma^i)_{\alpha\beta}$ are defined in Eq.(A11).

The supercurrents above are normalized in such a way that the standard commutation relation holds,

$$\{Q_\alpha\,\bar Q_{\dot\alpha}\} \;=\; 2\,P_{\alpha\dot\alpha},$$

and the central extension of the superalgebra then is

$$\{Q_\alpha\,Q_\beta\} \;=\; \frac{-iN}{8\pi^2}(\vec\sigma)_{\alpha\beta}\int d^3x\vec\nabla\left(\bar\lambda^a_{\dot\alpha}\bar\lambda^{a\dot\alpha}\right) \tag{A42}$$

modulo full superderivatives. Finally, we give (for future applications) a few expressions for SQCD (i.e. supersymmetric gluodynamics plus matter).

Consider for simplicity $SU(2)$ gauge group and one matter field in the fundamental representation (two subflavors). The corresponding chiral field carries the color and subflavor indices. The Lagrangian is

$$\mathcal{L}_M = \frac{1}{4}\int d^4\theta\left\{\bar S^{\alpha f}e^V S^{\alpha f}\right\} + \left(\frac{m}{4}\int d^2\theta\,S^{\alpha f}S_{\alpha f} \;+\; \mathrm{H.C.}\right) \tag{A43}$$

In components it has the form

$$\begin{aligned}
\mathcal{L} = \frac{1}{g^2}&\left[-\frac{1}{4}G^a_{\mu\nu}G^a_{\mu\nu} + \lambda^{\alpha,a}i\mathcal{D}_{\alpha\dot\alpha}\bar\lambda^{\dot\alpha,a} + \frac{1}{2}D^a D^a\right] + \\
& \psi^{f\alpha}i\mathcal{D}_{\alpha\dot\alpha}\bar\psi^{f\dot\alpha} + (\mathcal{D}_\mu\phi^{+f})(\mathcal{D}_\mu\phi^f) + F^{+f}F^f + \\
& i\sqrt{2}(\phi_1^+\lambda\psi_1 + \phi_1\bar\lambda\bar\psi_1 + \phi_2^+\lambda\psi_2 + \phi_2\bar\lambda\bar\psi_2) + \\
& \frac{1}{2}D^a\,(\phi_1^+T^a\phi_1 + \phi_2^+T^a\phi_2) + \\
& m\,(\phi_1 F_2 + \phi_1^+ F_2^+ + \phi_2 F_1 + \phi_2^+ F_1^+ + \psi_1\psi_2 + \bar\psi_1\bar\psi_2)
\end{aligned} \tag{A44}$$

The supercurrent is now modified,

$$J_{\alpha\beta\dot\beta} = 2\left\{\frac{1}{g^2}\left(iG^a_{\beta\alpha}\bar\lambda^a_{\dot\beta} - 3\epsilon_{\beta\alpha}D^a\bar\lambda^a_{\dot\beta}\right) + \right.$$

$$\left.\sqrt{2}\left[(\partial_{\alpha\dot\beta}\phi^+)\psi_\beta - i\,\epsilon_{\beta\alpha}F\bar\psi_{\dot\beta}\right] - \frac{\sqrt{2}}{6}\left[\partial_{\alpha\dot\beta}(\psi_\beta\phi^+) + \partial_{\beta\dot\beta}(\psi_\alpha\phi^+) - 3\epsilon_{\beta\alpha}\partial^\gamma_{\dot\beta}(\psi_\gamma\phi^+)\right]\right\}, \quad (A45)$$

and the central extension is

$$\{\bar Q_{\dot\alpha}\bar Q_{\dot\beta}\} = (-4i)(\vec\sigma)_{\dot\alpha\dot\beta}\int d^3x\vec\nabla\left\{\left[\mathcal{W} - \frac{N_c - N_f}{16\pi^2}\mathrm{Tr}W^2\right]\right\}_{\theta=0,\bar\theta=0}. \quad (A46)$$

It is also useful to write down the Konishi anomaly equation in the superfield formalism:

$$\bar D^2\left(\bar S^{\alpha f}\,e^V\,S^{\alpha f}\right) = 4m\,S^{\alpha f}S_{\alpha f} + \frac{1}{2\pi^2}\,\mathrm{Tr}W^2. \quad (A47)$$

For future references, let us also give the expression for the Lagrangian of the SQCD with the arbitrary gauge group and superpotential $\mathcal{W}$

$$\mathcal{L} = \left\{\frac{1}{4g_0^2}\mathrm{Tr}\int d^2\theta\,W^2 + H.c.\right\} + \frac{1}{4}\int d^4\theta\,\sum_i \bar Q_i e^V Q_i + \left\{\frac{1}{2}\int d^2\theta\,\mathcal{W}(Q_i) + H.c.\right\}. \quad (A48)$$

### APPENDIX B:

To obtain the three-dimensional Lagrangian we substitute expansion (123) into Lagrangian (120, 136) and integrate over $z$. We will need to compute the following set of integrals (brackets stay for the integration over $z$),

$$\langle\,a_i\,a_j\,a_k\,\rangle,\quad \langle\,a_i\,b_j\,b_k\,\rangle,\quad \langle\,\phi_w a_i\,a_j\,a_k\,\rangle,$$
$$\langle\,\phi_w a_i\,b_j\,b_k\,\rangle,\quad \langle\,a_i\,a_j\,a_k\,a_l\rangle,\quad \langle\,a_i\,a_j\,b_k\,b_l\,\rangle\ldots \quad (B1)$$

where $a_i$ and $b_i$ are the eigenmodes (124) and $\phi_w$ is the classical solution. Fortunately, to establish the existence of SUSY we do not have to compute the integrals, but rather to establish certain relations between them. With the aid of the completeness relation

$$\sum_n \lambda_n\,a_n(z)\,a_n(y) = \delta(z-y) \quad (B2)$$

where $\lambda_n$ is the normalization factor, and repeated use of Eq.(124) all the integrals can be reduced to the following two:

$$D_{ijk} = \sqrt{\kappa_i \lambda_j \lambda_k} \langle\, b_i\, a_j\, a_k\, \rangle; \quad A_{ijk} = \sqrt{\kappa_i \kappa_j \kappa_k} \langle\, b_i\, b_j\, b_k\, \rangle. \tag{B3}$$

Coefficient $A_{ijk}$ is symmetric, while $D_{ijk}$ is symmetric with respect to the last two indices. Note also, that since the operator $\mathcal{O}^1 \mathcal{O}^2(z)$ does not have a zero mode, then

$$D_{000} = A_{000} = 0. \tag{B4}$$

With the aid of the following redefinitions:

$$\alpha_n = \frac{\Psi_n^+ + i\Psi_n^-}{\sqrt{2}}; \quad \beta_n = \frac{\Psi_n^+ - i\Psi_n^-}{\sqrt{2}},$$

$$\tilde{a}_n = \frac{a_n + b_n}{\sqrt{2}}; \quad \tilde{b}_n = \frac{a_n - b_n}{\sqrt{2}},$$

$$B_{ijk} = D_{ijk} + D_{jik} + D_{kij} - A_{ijk},$$

$$C_{ijk} = D_{ijk} - D_{jik} - D_{kij} - A_{ijk}. \tag{B5}$$

the Lagrangian (120,136) can be seen as a component expansion of supersymmetric Lagrangian (133).

### APPENDIX C: PROPERTIES OF THE OPERATORS $\mathcal{O}^2\mathcal{O}^1$ AND $\mathcal{O}^1\mathcal{O}^2$

The operators $\mathcal{O}^2\mathcal{O}^1(z)$, $\mathcal{O}^1\mathcal{O}^2(z)$ have the following form:

$$\mathcal{O}^1\mathcal{O}^2(z) = -\partial_z^2 + \mathcal{W}''^2(\phi_w) - \mathcal{W}'\mathcal{W}'''(\phi_w),$$

$$\mathcal{O}^2\mathcal{O}^1(z) = -\partial_z^2 + \mathcal{W}''^2(\phi_w) + \mathcal{W}'\mathcal{W}'''(\phi_w). \tag{C1}$$

After we substitute the classical solution

$$\phi_w = \frac{m}{\lambda}\tanh(mz), \tag{C2}$$

we arrive at

$$\mathcal{O}^1\mathcal{O}^2(z) = -\partial_z^2 + 4m^2 - \frac{2m^2}{\cosh^2(mz)},$$

$$\mathcal{O}^2\mathcal{O}^1(z) = -\partial_z^2 + 4m^2 - \frac{6m^2}{\cosh^2(mz)}. \tag{C3}$$

Operator $\mathcal{O}^2\mathcal{O}^1(z)$ has two discrete eigenvalues and continuum,

$$\lambda_0^+ = 0; \quad \lambda_1^+ = \sqrt{3}m; \quad \lambda_k^+ = \sqrt{k^2 + 4m^2}. \tag{C4}$$

Operator $\mathcal{O}^1\mathcal{O}^2(z)$ has only one discrete mode and continuum,

$$\lambda_1^- = \sqrt{3}m; \quad \lambda_k^- = \sqrt{k^2 + 4m^2}. \tag{C5}$$

It is obvious that if $b_n$ is an eigenstate of $\mathcal{O}^1\mathcal{O}^2(z)$ then $a_n = \mathcal{O}^2 b_n$ is an eigenstate of $\mathcal{O}^2\mathcal{O}^1(z)$ with the same eigenvalue. Thus, two operators have exactly the same spectra, except for the zero mode of $\mathcal{O}^2\mathcal{O}^1(z)$. This is the type of symmetry one encounters in supersymmetric Quantum Mechanics [38].

---